\documentclass[aps,prd,twocolumn,showpacs,notitlepage,eqsecnum,
superscriptaddress,nofootinbib]{revtex4-1}

\usepackage[centertags]{amsmath} \usepackage{amssymb} \usepackage{latexsym}
\usepackage{enumerate} \usepackage{graphicx} \usepackage{mathrsfs}
\usepackage[colorlinks]{hyperref} \usepackage{stmaryrd} \usepackage{import}
\usepackage{tensor} \usepackage[usenames,dvipsnames]{xcolor} \usepackage{bm}
\usepackage{multirow} \usepackage{breakurl} \usepackage{float}
\usepackage{verbatim} \usepackage{mathrsfs}

\allowdisplaybreaks[1]

\definecolor{CiteColor}{rgb}{0,0.5,0} \hypersetup{citecolor=CiteColor}
\definecolor{RefColor}{rgb}{0.55,0,0} \hypersetup{linkcolor=RefColor}
\definecolor{darkgreen}{rgb}{0.2,0.7,0.2}

\newcommand{\bi}{\begin{itemize}} \newcommand{\ei}{\end{itemize}}
\newcommand{\be}{\begin{equation}} \newcommand{\ee}{\end{equation}}

\begin{document}

\title{Inspirals into a charged black hole}

\author{Ruomin Zhu}
\affiliation{Department of Physics and Astronomy, 
Oxford College of Emory University, Oxford, Georgia 30054, USA}
\author{Thomas Osburn}
\email{tosburn@emory.edu}
\affiliation{Department of Physics and Astronomy, 
Oxford College of Emory University, Oxford, Georgia 30054, USA}
\affiliation{Department of Physics and Astronomy, 
University of North Carolina, Chapel Hill, North Carolina 27599, USA}

\begin{abstract}
We model the quasicircular inspiral of a compact object into a more massive charged black hole. Extreme and intermediate mass-ratio inspirals are considered through a small mass-ratio approximation. Reissner-Nordstr$\ddot{\text{o}}$m spacetime is used to describe the charged black hole.  The effect of radiation reaction on the smaller body is quantified through calculation of electromagnetic and gravitational energy fluxes via solution of Einstein's and Maxwell's equations. Inspiral trajectories are determined by matching the orbital energy decay rate to the rate of radiative energy dissipation. We observe that inspirals into a charged black hole evolve more rapidly than comparable inspirals into a neutral black hole. Through analysis of a variety of inspiral configurations, we conclude that electric charge is an important effect concerning gravitational wave observations when the charge exceeds the threshold $|Q|/M \gtrsim 0.071 \sqrt{\epsilon}$, where $\epsilon$ is the mass ratio.
\end{abstract}

\maketitle

\section{Introduction}

Gravitational wave observations of compact binary systems have triggered a new era of astronomy. Successful Advanced LIGO \cite{aLIGO} and Advanced Virgo \cite{aVIRGO} observations of compact binary mergers \cite{LIGO1,LIGO2,LIGO3,LIGO4,LIGO5,LIGO6} have already solidified the foundations of gravitational physics~\cite{LIGOgrTest} and informed studies of compact object formation and evolution \cite{LIGOapj,LIGOns}. Over the next decade, ground-based gravitational wave detection rates will rise sharply through sensitivity enhancements \cite{LIGOsensitive}, establishment of the KAGRA \cite{KAGRA} detector, and the likely construction of LIGO-India \cite{indigo}. Ground-based gravitational wave detectors are sensitive in the broadband frequency range $10\,\text{Hz} \lesssim \nu \lesssim 10^4\,\text{Hz}$, which is suitable for observation of compact binary systems with total mass $1\, M_\odot \lesssim M+\mu \lesssim 10^3\, M_\odot$ \cite{LIGOfreq}, where $M$ and $\mu$ are the masses of each binary component. Ground-based gravitational wave detectors will be complemented strongly by the space-based LISA detector \cite{LISA} to be launched in 2034 as the European Space Agency's L3 mission. LISA will be sensitive in the broadband frequency range $10^{-5}\,\text{Hz} \lesssim \nu \lesssim 1\,\text{Hz}$, which is suitable for observation of compact binary systems with total mass $100\, M_\odot \lesssim M+\mu \lesssim 10^7\, M_\odot$ \cite{LISAfreq}. Finally, pulsar timing arrays such as IPTA \cite{PTA} probe the lowest gravitational wave frequencies in the range $10^{-9}\,\text{Hz} \lesssim \nu \lesssim 10^{-6}\,\text{Hz}$, which is suitable for observation of compact binary systems with total mass $10^8\, M_\odot \lesssim M+\mu \lesssim 10^{10}\, M_\odot$ \cite{IPTA}. 

Regardless of the detection scheme, theoretical models are instrumental in gravitational wave detection via matched filtering. Theoretical models are also needed to determine the parameters describing gravitational wave sources. Compact binary systems are described by a host of parameters including the mass of each binary component, the spin of each binary component, and the orbital eccentricity, separation, and inclination \cite{BaraCutl04}. One potential property that is often disregarded in theoretical models of compact objects is electric charge. Indeed, it is straightforward to argue that any excess electric charge would be neutralized rapidly in realistic scenarios \cite{Nara05}. However, mechanisms predicting the existence of charged compact objects have been proposed through classical arguments \cite{Wald74,Cohe75,Lee01} and more exotic arguments such as those involving dark matter \cite{RujuETC90,SiguETC04,CardETC16}. Furthermore, even if currently proposed charge explanations were improbable, it would still be useful to investigate the existence of charged compact objects in case of a charging mechanism that is yet undiscovered. Therefore, the purpose of this paper is to quantify how strongly the gravitational and electromagnetic dynamics of compact binary systems are affected by introducing electric charge to one of the binary components.

The above scenario is a version of the relativistic two-body problem, with solutions accessible through Einstein's and Maxwell's equations (in curved spacetime). Exact solutions of the relativistic two-body problem are not generally known, but depending on the properties of the system, various approximation schemes are available. When the binary components have comparable masses and a small separation, no analytic approximation schemes apply. Instead, numerical approximations are made to solve the nonlinear Einstein-Maxwell equations on a supercomputer. This scheme is called numerical relativity. Numerical relativity has been used to simulate head-on collisions of charged black holes \cite{Zilh12,Zilh14}, but little else has been done with numerical relativity regarding charged compact binary systems. Also in the comparable mass and small separation regime, analytic kludges have been used to develop charged compact binary models that predict electromagnetic radiation from nearby accelerated particles \cite{Zhan16}. Numerical relativity is not well suited to simulate compact binary systems when the masses are highly dissimilar or the separation is large (due to catastrophic separation of timescales), but alternate schemes apply in those cases.

When the binary components have a large separation, post-Newtonian theory describes compact binary dynamics through an expansion in powers of the small velocity. Some studies considering charged compact binaries have applied post-Newtonian theory \cite{Bini16}, but very little has concerned gravitational wave astronomy. When the binary component masses differ significantly, an expansion in powers of the small mass ratio, $\epsilon\equiv \mu/M$, called black hole perturbation theory (BHPT), is applicable. An advantage of BHPT is that it imposes no restrictions on the velocity or gravitational field strength while maintaining validity in the small mass-ratio regime ($\epsilon\ll 1$). This work applies BHPT to systems where the more massive binary component (with mass $M$) is charged and the less massive binary component (with mass $\mu$) is neutral. Because there is an upper limit on neutron star masses, the larger binary component must be a black hole when $\epsilon \ll 1$ (if $\mu \gtrsim 1 \, M_\odot$). For simplicity we assume that the charged black hole does not have spin. Such black holes are described by Reissner-Nordstr$\ddot{\text{o}}$m (RN) spacetime. Note that some models explaining the presence of electric charge require a spinning black hole, and for those cases the analysis of this work would be qualitative in nature.

A number of studies have applied BHPT to RN spacetime. Zerilli was the first to derive the equations describing gravitational and electromagnetic perturbations of RN black holes \cite{Zeri74}. Zerilli's equations were then solved to model radiation from the radial infall of a compact object into a charged black hole \cite{John73,John74}. Recent work has reproduced those results to quantify how electric charge affects gravitational wave emission from radial plunges \cite{CardETC16}. Moncrief independently studied whether RN spacetime is stable under perturbations \cite{Monc74a,Monc74b,Monc75}, and that work was generalized to consider scattering by and quasinormal modes of RN black holes \cite{Chan79,Gunt80}. BHPT has also been applied to describe how RN spacetime is perturbed by a static particle \cite{Bini07}. Radiation reaction on the smaller binary component occurs through a mechanism called the self-force  \cite{MinoSasaTana97,QuinWald97,DetwWhit03,Pois11}. Self-force studies in charged black hole spacetimes have been developed through scalar field toy models \cite{Bara00,Burk01} and the groundwork is being laid for more realistic charged self-force scenarios \cite{ZimmPois14,Linz14,Zimm15}. Leading-order self-force effects are accessible through an adiabatic approximation \cite{Poun05}. Adiabatic approximations are equivalent to time averaging the self-force. An advantage of adiabatic approximations is that they are accessible through dissipative flux calculations, which converge rapidly and avoid the complexities of local regularization. Higher order effects, such as the conservative part of the self-force, enter through the postadiabatic expansion \cite{Hind08}. This work uses an adiabatic approximation to model the quasicircular inspiral of a point mass into a RN black hole. 

The novel scientific achievements of this work are summarized as follows. The master function formalism originally developed by Moncrief \cite{Monc74a,Monc74b,Monc75} to simplify the RN perturbation equations is generalized to the inhomogeneous case (see Appendixes \ref{sec:decomp} and \ref{sec:dipole}). Using this generalized formalism, we describe numerically the gravitational and electromagnetic radiation from a compact object in a circular orbit around a RN black hole (see Sec. \ref{sec:master}). From our numerical calculations we quantify the rate at which orbital energy is dissipated by radiation for arbitrary electric charge and orbital radius (see Sec. \ref{sec:insp}). Armed with the energy flux as a function of orbital radius, we consider radiation reaction on the smaller binary component through adiabatic and quasi-circular approximations (see Sec. \ref{sec:insp}). Our model is the first to describe the inspiral of a small compact object into an arbitrarily charged RN black hole to leading order in the postadiabatic expansion. Potential observations of electromagnetic radiation from this type of system are considered briefly (see Sec. \ref{sec:results}). Finally, our inspiral model is applied to quantify the level at which electric charge could affect gravitational wave observations (see Sec. \ref{sec:results}).

Sections \ref{sec:pert} and \ref{sec:orb} and Appendix \ref{sec:tsh} establish the theoretical background that serves as the foundation for our analysis. Section \ref{sec:alg} explains our numerical algorithm and quantifies computational cost and accuracy. Finally, Sec. \ref{sec:conc} summarizes key results and investigates future directions of inquiry related to charged compact binary systems. Throughout this paper we adopt units such that $c = G = 1$, $\mu_0 = \varepsilon_0^{-1} = 4\pi$, and we use the metric signature $(-+++)$.

\section{Perturbations of a charged black hole}
\label{sec:pert}

We model the charged binary system using first-order perturbations of the Einstein-Maxwell equations in RN spacetime. In this scheme the mass ratio, $\epsilon\equiv\mu/M$, is used as a small parameter to expand the spacetime metric, $g_{\alpha\beta}$, and the electromagnetic potential four-vector, $A_\alpha$,
\begin{align}
&g_{\alpha\beta} = g^{(0)}_{\alpha\beta} + g^{(1)}_{\alpha\beta} + \mathcal{O}(\epsilon^2) ,
\\
&A_\alpha = A^{(0)}_\alpha + A^{(1)}_\alpha + \mathcal{O}(\epsilon^2) .
\end{align}
$A_\alpha$ and $g_{\alpha\beta}$ are governed by the Einstein-Maxwell equations
\begin{align}
\label{eq:EinMax}
\nabla_\beta F^{\alpha\beta} = 4\pi J^\alpha , \qquad \qquad G_{\alpha\beta} = 8\pi T_{\alpha\beta} ,
\end{align}
where $F^{\alpha\beta}$ is the electromagnetic field tensor
\begin{align}
F_{\alpha\beta} = \nabla_\alpha A_\beta - \nabla_\beta A_\alpha, 
\end{align}
$J^\alpha$ is the current density four-vector, $G_{\alpha\beta}$ is the Einstein tensor, and $T_{\alpha\beta}$ is the stress-energy tensor
\begin{align}
&T^{\alpha\beta} = \mu \frac{u^\alpha u^\beta}{u^t r_p^2 \sin{\theta_p}} \delta(r-r_p)\delta(\theta-\theta_p)\delta(\varphi-\varphi_p) \notag
\\&\qquad\qquad\qquad + \frac{1}{4\pi}\left( F^{\alpha}_{\;\;\gamma} F^{\beta\gamma} - \frac{1}{4}g^{\alpha\beta}F_{\gamma\nu}F^{\gamma\nu} \right).
\label{eq:Tab}
\end{align}
Here $\delta$ is the Dirac delta function, $r_p$, $\theta_p$, and $\varphi_p$ denote the time-dependent position of the small compact object, and $u^\alpha$ is its four-velocity. The first term in Eq.~\eqref{eq:Tab} represents the stress energy of a point mass, and the last term in Eq.~\eqref{eq:Tab} is the electromagnetic stress energy. The Einstein tensor, $G_{\alpha\beta}$, is generated by applying a nonlinear second-order differential operator to $g_{\alpha\beta}$. The electromagnetic and gravitational fields are coupled. The presence of electromagnetic terms in the stress-energy tensor introduces $A_{\alpha}$ into the gravitational field equations, and the presence of covariant derivatives ($\nabla_\alpha$) introduces $g_{\alpha\beta}$ into the electromagnetic field equations.

The leading-order terms in the small mass-ratio expansion, $g^{(0)}_{\alpha\beta}$ and $A^{(0)}_\alpha$, are exact solutions of Eq. \eqref{eq:EinMax} describing the larger binary component. Here $g^{(0)}_{\alpha\beta}$ is the RN metric (we adopt Boyer-Lindquist coordinates),
\begin{align}
g^{(0)}_{tt} = -f , \qquad & g^{(0)}_{rr} = \frac{1}{f}, \qquad g^{(0)}_{\theta\theta} = \frac{g^{(0)}_{\varphi\varphi}}{\sin^2\theta} = r^2 , 
\\
f & \equiv 1-\frac{2M}{r}+\frac{Q^2}{r^2} .
\end{align}
All off-diagonal components of the RN metric vanish ($g^{(0)}_{\alpha\beta}=0$ when $\alpha \ne \beta$). The leading term in the electromagnetic expansion, $A^{(0)}_\alpha$, is a vacuum solution of Maxwell's equations compatible with the RN metric,
\begin{align}
A^{(0)}_t &= \frac{Q}{r}, \qquad \qquad A^{(0)}_r = A^{(0)}_\theta = A^{(0)}_\varphi = 0 .
\end{align}
The first-order gravitational and electromagnetic perturbations, $g^{(1)}_{\alpha\beta}$ and $A^{(1)}_\alpha$, are determined by expanding Eq. \eqref{eq:EinMax} through linear order in $\epsilon$. We specialize to the case where the smaller binary component has no electric charge by requiring that $J^\alpha=0$. 

The spherical symmetry of RN spacetime admits a tensor spherical harmonic decomposition of the electromagnetic and gravitational perturbations. For each spherical harmonic ($l$,$m$) mode there are four radial functions describing the electromagnetic perturbations and ten radial functions describing the gravitational perturbations. The field equations conveniently decouple into two sets, odd-parity and even-parity, based on how the tensor spherical harmonics behave under a parity transformation. For equatorial source motion, the odd-parity source terms vanish when $l+m$ is an even number, and the even-parity source terms vanish when $l+m$ is an odd number. In this scenario it is natural to consider only the odd-parity field equations when $l+m$ is odd and only the even-parity field equations when $l+m$ is even. For full details see Appendix \ref{sec:tsh}.

\section{Reissner-Nordstr$\ddot{\text{O}}$m orbital motion}
\label{sec:orb}

In our perturbative scheme, the stress energy of the smaller binary component sources the first-order fields. Description of this source mechanism requires knowledge of the orbital characteristics of the small compact object. Ignoring radiation reaction effects, point masses follow timelike geodesics of the background spacetime. Even when radiation reaction is included, the trajectory will continue to resemble a geodesic during short time intervals ($\Delta t \sim M$). Inclusion of radiation reaction in the source motion affects the system at a higher order than we are considering. Therefore, our model uses geodesic motion to describe the source terms present in the field equations.

Both circular and eccentric geodesics will be important to consider \cite{Hopm05}, but this work focuses solely on circular orbital motion. Chandrasekhar's textbook \cite{Chan92} is a useful reference for studying this section in detail. For circular motion the components of the position vector are $r_p=$ const, $\theta_p = \pi/2$, and $\varphi_p = \Omega\, t$, where $\Omega$ is the angular speed. Functions of $r$ with a $p$ subscript are assumed to be evaluated at $r=r_p$ (a useful example is $f_p = 1-2M/r_p+Q^2/r_p^2$). The orbital specific energy, $\mathcal{E}$, and specific angular momentum, $\mathcal{L}$, follow from symmetries of RN spacetime,
\begin{align}
\mathcal{E} &= \frac{r_p f_p}{\sqrt{r_p^2-3r_pM+2Q^2}},
\label{eq:orben}
\\
\mathcal{L} &= r_p\sqrt{\frac{r_p M-Q^2}{r_p^2-3r_p M +2 Q^2}} .
\end{align}
It is useful to express $u^\alpha$ in terms of the orbital energy and angular momentum,
\begin{align}
u^t = \mathcal{E}/f_p, \qquad\;\; u^\varphi = \mathcal{L}/r_p^2, \qquad\;\; u^r=u^\theta = 0.
\end{align}
The angular speed is the derivative of $\varphi_p$ with respect to $t$,
\begin{align}
\label{eq:Omega}
\Omega &= \frac{d\varphi_p}{dt} = \frac{u^\varphi}{u^t} = \frac{\sqrt{r_p M-Q^2}}{r_p^2} .
\end{align}
For each valid $M$ and $Q$ there exists an orbital radius below which circular motion is unstable. This orbit is referred to as the innermost stable circular orbit (ISCO), and its radius, $r_\text{ISCO}$, satisfies a cubic equation with a single real root,
\begin{align}
M r_\text{ISCO}^3 - 6 M^2 r_\text{ISCO}^2 + 9 M Q^2 r_\text{ISCO} - 4Q^4 = 0.
\end{align}
Under the effect of radiation reaction, the orbit will plunge rapidly when $r_p<r_\text{ISCO}$.

The angular speed has special importance in the field equations through a mechanism involving the spherical harmonic source decomposition. Orthogonality of our angular basis is used to decompose the stress energy of the point mass into tensor spherical harmonic modes. This process involves integrating products of spherical harmonic complex conjugates and Dirac delta functions, including $\delta(\varphi-\Omega t)$. Through this process the time-domain source terms gain overall factors of $e^{-im\Omega t}$. Periodicity of the source implies periodicity of the solution to the linear inhomogeneous field equations. The periodic nature of the solution is amenable to analysis in the frequency domain via Fourier series. In this scenario the frequency domain is accessed through the replacement $\partial_t \rightarrow -i\, \omega_m$, with $\omega_m \equiv m \Omega$. See Appendix \ref{sec:tsh} for more details.

\section{Master equations}
\label{sec:master}

For each spherical harmonic ($l$, $m$) mode, the Einstein-Maxwell equations reduce to a system of coupled ordinary differential equations (ODEs) describing gravitational and electromagnetic radial functions. Even-parity modes involve ten coupled ODEs describing seven gravitational and three electromagnetic radial functions, while odd-parity modes involve four coupled ODEs describing three gravitational and one electromagnetic radial function. We reduce the number of even-parity radial functions from ten to six and similarly reduce the number of odd-parity radial functions from four to three through a suitable gauge choice (the Regge-Wheeler gauge). Zerilli \cite{Zeri74} and Moncrief \cite{Monc74a,Monc74b,Monc75} demonstrated that further simplifications are possible by introducing gravitational and electromagnetic ``master functions.'' Here we extend Moncrief's master function formalism to the inhomogeneous case (although some alternate mathematical choices are made in this work). For each even- or odd-parity mode, the radial functions associated with $g^{(1)}_{\alpha\beta}$ are constructed from a single gravitational master function, $h^\text{even}_{lm}(r)$ or $h^\text{odd}_{lm}(r)$, and the radial functions associated with $A^{(1)}_{\alpha}$ are constructed from a single electromagnetic master function, $a^\text{even}_{lm}(r)$ or $a^\text{odd}_{lm}(r)$. In order for the Einstein-Maxwell equations to be satisfied, the master functions must satisfy a simplified system of ODEs called the ``master equations.'' Because the even-parity master equations have the same general form as the odd-parity master equations, the even/odd superscripts are omitted in this section. For full details concerning the master equations see Appendix \ref{sec:decomp}. The dipole ($l=1$) modes require some special treatment; see Appendix \ref{sec:dipole}.

The master functions satisfy the following system of ODEs:
\begin{align}
\label{eq:master}
\left( \frac{d^2}{dr_*^2} + \omega_m^2 + \left[\begin{array}{cc} \alpha_{lm} & \beta_{lm} \\ \gamma_{lm} & \sigma_{lm} \end{array} \right] \right) \left[\begin{array}{c} h_{lm} \\ a_{lm} \end{array} \right] &= \left[\begin{array}{c} S_{lm} \\ Z_{lm} \end{array} \right] ,
\end{align}
where $r_*$ is the tortoise coordinate
\begin{align}
r_* &= r+ \frac{r_+^2}{r_+-r_-}\ln\left( \frac{r-r_+}{M} \right)- \frac{r_-^2}{r_+-r_-}\ln\left( \frac{r-r_-}{M} \right) .
\end{align}
The radial positions $r_\pm$ represent the inner ($-$) and outer ($+$) event horizons of the RN black hole,
\begin{align}
r_\pm = M \pm \sqrt{M^2-Q^2} .
\end{align}
Note that $\dfrac{dr_*}{dr} = f^{-1}$. Also note the limiting behavior of $r_*$,
\begin{align}
 \lim_{r\rightarrow r_+}r_* &= -\infty , \qquad\qquad \lim_{r\rightarrow \infty} r_* = +\infty.
\end{align}
The ODE coefficients, $\alpha_{lm}$, $\beta_{lm}$, $\gamma_{lm}$, $\sigma_{lm}$, and source terms, $S_{lm}$, $Z_{lm}$, are each functions of $r$. The ODE coefficients and source terms have different forms depending on the parity (even/odd) of the master function. One property shared by all ODE coefficients is that they vanish approaching $r=r_+$ and $r=\infty$,
\begin{align}
\underset{r_* \rightarrow \pm \infty\;\;\;\;\;\;}{\text{lim}\;\, \alpha_{lm}} = \underset{r_* \rightarrow \pm \infty\;\;\;\;\;\;}{\text{lim}\;\, \beta_{lm}} =\underset{r_* \rightarrow \pm \infty\;\;\;\;\;\;}{\text{lim}\;\, \gamma_{lm}} =\underset{r_* \rightarrow \pm \infty\;\;\;\;\;\;}{\text{lim}\;\, \sigma_{lm}} = \; 0.
\end{align}
This property, in the context of Eq. \eqref{eq:master}, requires that $h_{lm}$ and $a_{lm}$ behave as traveling waves in the near horizon zone ($r-r_+\ll M$) and wave zone ($r \gg |\omega_m|^{-1}$) with a wavelength (measured according to $r_*$) of $2\pi/|\omega_m|$. The $r$ dependence of the ODE sources involves Dirac delta functions
\begin{align}
\label{eq:source}
\left[\begin{array}{c} S_{lm}(r) \\ Z_{lm}(r) \end{array} \right] = \left[\begin{array}{c} B_{lm} \\ D_{lm} \end{array} \right]\delta(r-r_p) + \left[\begin{array}{c} F_{lm} \\ H_{lm} \end{array} \right] \delta'(r-r_p),
\end{align}
where $B_{lm}$, $D_{lm}$, $F_{lm}$, and $H_{lm}$ are constants determined by the orbital characteristics, and a prime denotes differentiation with respect to $r$. Note that the $\theta$ and $\varphi$ Dirac delta functions present in Eq. \eqref{eq:Tab} have been eliminated from Eq. \eqref{eq:source} through decomposition of the point mass stress energy into spherical harmonic modes. Derivatives of the $r$ Dirac delta functions appear because the field equations are differentiated in our master function formalism.

Equation \eqref{eq:master} has four independent homogeneous solutions. We denote each independent homogeneous solution with a superscript that implies certain boundary behavior. The two ``outgoing'' homogeneous solutions propagate toward $r_*=+\infty$ when $r\gg |\omega_m|^{-1}$,
\begin{align}
\label{eq:out}
\left[ \begin{array}{c} h_{lm}^{0+} \\ a_{lm}^{0+} \end{array} \right] \simeq e^{+i\omega_m r_*} \left[ \begin{array}{c} 1 \\ 0 \end{array} \right] , \;\;\;\;\; \left[ \begin{array}{c} h_{lm}^{1+} \\ a_{lm}^{1+} \end{array} \right] \simeq e^{+i\omega_m r_*} \left[ \begin{array}{c} 0 \\ 1 \end{array} \right] ,
\end{align}
The two ``downgoing'' homogeneous solutions propagate toward $r_*=-\infty$ when $r-r_+ \ll M$,
\begin{align}
\label{eq:down}
\left[ \begin{array}{c} h_{lm}^{0-} \\ a_{lm}^{0-} \end{array} \right] \simeq e^{-i\omega_m r_*} \left[ \begin{array}{c} 1 \\ 0 \end{array} \right] , \;\;\;\;\; \left[ \begin{array}{c} h_{lm}^{1-} \\ a_{lm}^{1-} \end{array} \right] \simeq e^{-i\omega_m r_*} \left[ \begin{array}{c} 0 \\ 1 \end{array} \right] .
\end{align}
This outgoing and downgoing set of homogeneous solutions is not unique. We could form a new basis of homogeneous solutions through linear combinations of Eqs. \eqref{eq:out} and \eqref{eq:down}. However, our chosen basis is convenient for finding the inhomogeneous solution representing wave propagation away from the source (the retarded solution). We expand each homogeneous solution in a power series near the boundary (either $r\simeq \infty$ or $r\simeq r_+$)  to generate initial values for numerical integration of Eq. \eqref{eq:master}. These numerical integrations determine the global homogeneous solutions. See Appendix \ref{sec:boundary} for full details of the boundary expansions.

Because the source terms involve Dirac delta functions with no more than one derivative, the inhomogeneous solution can be expressed as a piecewise function of homogeneous solutions,
\begin{align}
\label{eq:inhomo}
&\left[\begin{array}{c} h_{lm} \\ a_{lm} \end{array} \right] = \left( C_{lm}^{0+} \left[ \begin{array}{c} h_{lm}^{0+} \\ a_{lm}^{0+} \end{array} \right] + C_{lm}^{1+} \left[ \begin{array}{c} h_{lm}^{1+} \\ a_{lm}^{1+} \end{array} \right]  \right) \Theta(r-r_p) \notag
\\&\qquad\;\;\; +\left( C_{lm}^{0-} \left[ \begin{array}{c} h_{lm}^{0-} \\ a_{lm}^{0-} \end{array} \right] + C_{lm}^{1-} \left[ \begin{array}{c} h_{lm}^{1-} \\ a_{lm}^{1-} \end{array} \right]  \right) \Theta(r_p-r) ,
\end{align}
where $\Theta$ is the Heaviside step function. Our specific basis of homogeneous solutions was chosen so that Eq. \eqref{eq:inhomo} represents the retarded solution of the field equations. The normalization coefficients, $C_{lm}^{j\pm}$, are determined by requiring that the Dirac delta functions (and derivatives) vanish when Eq. \eqref{eq:inhomo} is substituted into Eq. \eqref{eq:master} (recall that the the Dirac delta function is the derivative of the Heaviside step function). This is equivalent to applying variation of parameters by integrating Green's function against the ODE sources. The result of this procedure is the following linear system (involving the Wronskian matrix) that is satisfied by the normalization coefficients:
\begin{align}
&\left[\begin{array}{cccc} h_{lm}^{0+} & h_{lm}^{1+} & h_{lm}^{0-} & h_{lm}^{1-} 
\\ a_{lm}^{0+} & a_{lm}^{1+} & a_{lm}^{0-} & a_{lm}^{1-} 
\\ \partial_{r_*} h_{lm}^{0+} & \partial_{r_*} h_{lm}^{1+} & \partial_{r_*} h_{lm}^{0-} & \partial_{r_*} h_{lm}^{1-}
\\ \partial_{r_*} a_{lm}^{0+} & \partial_{r_*} a_{lm}^{1+} & \partial_{r_*} a_{lm}^{0-} & \partial_{r_*} a_{lm}^{1-}  \end{array} \right]_p \left[\begin{array}{c} C_{lm}^{0+} \\ C_{lm}^{1+} \\ -C_{lm}^{0-} \\ -C_{lm}^{1-} \end{array} \right] \notag
\\&\qquad\qquad\;\;\; = \frac{1}{r_p^3 f_p^2} \left[\begin{array}{c} r_p^3 F_{lm} \\ r_p^3 H_{lm} \\ r_p^3 f_p B_{lm}-2(Q^2-Mr_p) F_{lm} \\ r_p^3 f_p D_{lm}-2(Q^2-Mr_p) H_{lm} \end{array} \right] ,
\label{eq:wron}
\end{align}
where all $r$ dependent functions have been evaluated at $r=r_p$ (as implied by the $p$ subscripts), and $\partial_{r_*}$ represents an ordinary derivative with respect to $r_*$.

\section{Quasicircular inspirals}
\label{sec:insp}

Radiation reaction arises through the interaction of the small body with the gravitational and electromagnetic perturbations. This mechanism is called the self-force. For a thorough treatment of the self-force, see \cite{Pois11}. In this work we make a leading-order approximation equivalent to averaging the self-force called the adiabatic approximation. Furthermore, we assume that the motion of the small body is well approximated by a circular geodesic with a slowly changing radius throughout the inspiral. Under these adiabatic and quasicircular approximations, the inspiral dynamics are encoded by the rate of radiative energy dissipation.  By solving the Einstein-Maxwell equations we are able to calculate this energy dissipation rate.

After choosing an orbital radius, $r_p$, and finding the particular solution of Eq. \eqref{eq:master} for all spherical harmonic modes, the rate of radiative energy dissipation can be calculated. We refer to this average power measurement as the energy flux, $\langle \dot{E} \rangle$. In this work angle brackets, $\langle\rangle$, indicate a time ($t$) average and an overdot indicates a time ($t$) derivative. The total energy flux has four components: the gravitational flux propagating toward $r=\infty$, $\langle \dot{E}^+_\text{G} \rangle$, the electromagnetic flux propagating toward $r=\infty$, $\langle \dot{E}^+_\text{EM} \rangle$, the gravitational flux propagating toward the event horizon of the charged black hole, $\langle \dot{E}^-_\text{G} \rangle$, and the electromagnetic flux propagating toward the event horizon of the charged black hole, $\langle \dot{E}^-_\text{EM} \rangle$,
\begin{align}
\label{eq:Edot}
\langle \dot{E} \rangle &= \langle \dot{E}^+_\text{G} \rangle + \langle \dot{E}^+_\text{EM} \rangle + \langle \dot{E}^-_\text{G} \rangle + \langle \dot{E}^-_\text{EM} \rangle.
\end{align}
Each of these flux components is given by Poynting's theorem and its gravitational equivalent.

The $r$-component of the Poynting vector appears as the $tr$-component of the stress-energy tensor, $T^{tr}$. Integrating $T^{tr}$ over the 2-sphere at $r=\infty$ determines the outgoing electromagnetic energy flux, $\langle \dot{E}^+_\text{EM} \rangle$,
\begin{align}
\label{eq:infFlux}
\langle \dot{E}^+_\text{EM} \rangle  &= \bigg\langle \lim_{r\rightarrow \infty} \int T^{tr} r^2 d\Omega \bigg\rangle .
\end{align}
In this scenario, the stress-energy tensor is expanded through order $\epsilon^2$. The same procedure applies at the event horizon, although care must be taken to ensure that the surface integral involves a suitable proper area. Our chosen radial coordinate, $r$, gives the correct proper area at any radius $r\ge r_+$ (when using the naive area element $r^2 d\Omega$). Therefore, it is straightforward to generalize Eq. \eqref{eq:infFlux} to the horizon flux case
\begin{align}
\langle \dot{E}^-_\text{EM} \rangle  &= \bigg\langle \lim_{r\rightarrow r_+} \int T^{tr} r^2 d\Omega \bigg\rangle .
\end{align}
The only terms that survive the time and angle averaging processes are those involving the product of an ($l$, $m$) mode with an ($l$, $-m$) mode, resulting in the following equation for the electromagnetic energy fluxes in terms of the normalization constants,
\begin{align}
\label{eq:EMflux}
\langle \dot{E}^\pm_\text{EM} \rangle &= \sum_{l=1}^{\infty} \sum_{m=1}^l \frac{l(l+1)\omega_m^2 |C_{lm}^{1\pm}|^2}{2 \pi} .
\end{align}
To derive Eq. \eqref{eq:EMflux} we leveraged the fact that the coefficients $C_{l,m}^{j\pm}$ and $C_{l,-m}^{j\pm}$ are complex conjugates of each other. Note that we have not distinguished between the odd-parity, even-parity, or dipole master functions in Eq. \eqref{eq:EMflux}. The definitions of $a_{lm}^\text{odd}$ and $a_{lm}^\text{even}$ were chosen specifically to satisfy the same flux equations. 

Similar techniques apply to the gravitational energy flux. An effective gravitational stress-energy tensor can be constructed from the non-linear part of the Einstein tensor, $G_{\mu\nu}$. Through a similar averaging process the following relationship is derived for the gravitational energy fluxes
\begin{align}
\label{eq:Gflux}
\langle \dot{E}^\pm_\text{G} \rangle &= \sum_{l=2}^{\infty} \sum_{m=1}^l \frac{(l+2)(l+1)l(l-1)\omega_m^2 |C_{lm}^{0\pm}|^2}{32 \pi} .
\end{align}
The dependence of the electromagnetic energy flux on $C_{lm}^{1\pm}$ and the gravitational energy flux on $C_{lm}^{0\pm}$ is a consequence of our chosen basis of homogeneous solutions; see Eqs. \eqref{eq:out}-\eqref{eq:inhomo}.

The energy flux is used to drive the orbital evolution by enforcing the first law of thermodynamics. If energy is carried away by gravitational and electromagnetic waves, then the orbital energy must decrease accordingly,
\begin{align}
\mu \frac{d\mathcal{E}}{dt} &= - \langle \dot{E} \rangle .
\end{align}
The orbital specific energy, $\mathcal{E}$, depends on $r_p$, which we promote from a constant to a function of time to represent the inspiral. It is useful to cast the evolution equation in terms of $r_p(t)$ by analyzing Eq. \eqref{eq:orben}
\begin{align}
\frac{d r_p}{dt} &= -\dfrac{\langle \dot{E} \rangle}{\mu} \left( \dfrac{\partial \mathcal{E}}{\partial r_p} \right) ^{-1} , \notag 
\\&= -\dfrac{\langle \dot{E} \rangle}{\mu}\frac{2 r_p^2 \left(r_p^2-3 M r_p+2 Q^2\right)^{3/2}}{\left(M r_p  (r_p^2-6 M r_p+9 Q^2 )-4 Q^4\right)} .
\label{eq:inspr}
\end{align}
The azimuth of the small body, $\varphi_p(t)$, is calculated by integrating the slowly evolving angular speed given in Eq. \eqref{eq:Omega}
\begin{align}
\frac{d\varphi_p}{dt} &= \Omega(t) = \frac{\sqrt{r_p M-Q^2}}{r_p^2} .
\label{eq:inspphi}
\end{align}
Equations \eqref{eq:inspr} and \eqref{eq:inspphi} form a system of ODEs that, when solved numerically, approximate the position of the inspiraling small body to leading order in the postadiabatic expansion. One strategy is to couple Eq. \eqref{eq:inspr} directly to Eqs. \eqref{eq:EMflux} and \eqref{eq:Gflux} by re-solving the field equations at each integration step during the orbital evolution. One downside of that strategy is, when a large number of different orbital integrations are performed, the same field equations would often be re-solved at the same radii during different inspiral evolutions. Considering that solving the field equations is the most time-consuming step numerically, it would be advantageous to avoid redundancy in that area. Our strategy is to presolve the field equations for a large variety of $Q$ and $r_p$ values, and then interpolate $\langle \dot{E} \rangle$ as a function of $r_p$ for each $Q$ (the $\epsilon$ dependence of $\langle \dot{E} \rangle$ factors out). Then the interpolant of $\langle \dot{E} \rangle$ is coupled to Eq. \eqref{eq:inspr}, which avoids redundantly re-solving the field equations.

\section{Numerical algorithm}
\label{sec:alg}

\begin{figure}
\includegraphics[width=3.35in]{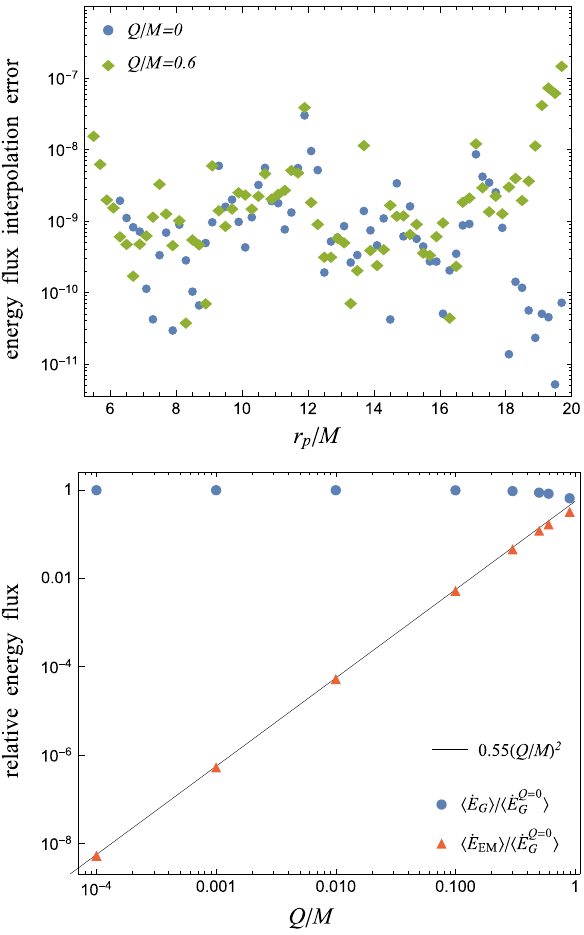}
\caption{\label{fig:intErrFlux}Top: The maximum relative error from energy flux interpolation is plotted as a function of orbital radius for two different charge values, $Q$. For this calculation we reduced the density of interpolation data by half to facilitate independent comparison of interpolant with unused flux data. This lower density interpolant has a maximum relative error of $10^{-7}$. For inspiral calculations we use the full density interpolant to further improve accuracy. Bottom: Gravitational and electromagnetic energy fluxes are plotted as a function of $Q$. The fluxes are scaled relative to the $Q=0$ gravitational flux. The orbital radius for this comparison is $r_p=7M$. We observe that the electromagnetic energy flux is proportional to $Q^2$ when $Q$ is small.}
\end{figure}

The numerical tools we employ include \textsc{Python} 3 (with \textsc{NumPy} and \textsc{SciPy}) and \emph{Mathematica}. \textsc{Python} is used to solve the field equations and calculate the energy flux while \emph{Mathematica} is used to evolve the inspiral. The following list details our numerical procedure:

\begin{enumerate}[(1)]
\item
A charge in the range $|Q|< M$ is chosen.
\item
An orbital radius in the range $r_p \ge r_\text{ISCO}$ is chosen.
\item
A tensor spherical harmonic ($l$, $m$) mode is chosen.
\begin{enumerate}
\item
$l$ is restricted to the range $l\ge 1$.
\item
$m$ is restricted to the range $1 \le m \le l$.
\item
If $l+m$ is even, we use the even-parity equations.
\item
If $l+m$ is odd, we use the odd-parity equations.
\end{enumerate}
\item
A custom \textsc{Python} function is used to generate initial values for numerical integration of the homogeneous solutions.
\begin{enumerate}[a.]
\item
Equation \eqref{eq:infExp} (with $j_\text{max}=30$) generates initial data for the solutions described by Eq. \eqref{eq:out}.
\item
Equation \eqref{eq:horExp} (with $j_\text{max}=30$) generates initial data for the solutions described by Eq. \eqref{eq:down}.
\item
Independent solutions are produced by selecting independent sets of starting coefficients in the recurrence relations; see Appendix \ref{sec:boundary}.
\end{enumerate}
\item
The homogeneous version of Eq. \eqref{eq:master} is integrated numerically using \emph{scipy.integrate.odeint} in \textsc{Python} (with accuracy tolerance = $10^{-13}$) for each set of initial values.
\begin{enumerate}[a.]
\item
The initial position $r_i = 30/|\omega_m| + 10M$ is used for the solutions described by Eq. \eqref{eq:out}.
\item
The initial position $r_i = r_+ + 10^{-8} M$ is used for the solutions described by Eq. \eqref{eq:down}.
\item
The final position $r_f = r_p$ is used for all homogeneous integrations.
\end{enumerate}
\item
The inhomogeneous solution is found using Eq. \eqref{eq:inhomo}.
\begin{enumerate}[a.]
\item
The Wronskian matrix is generated using the homogeneous solutions evaluated at $r=r_p$.
\item
The source vector is generated using the orbital characteristics implied by $Q$ and $r_p$.
\item
The normalization coefficients are calculated by solving Eq. \eqref{eq:wron} using \emph{numpy.linalg.solve} in \textsc{Python}.
\end{enumerate}
\item
Steps (3)-(6) are repeated for all $l$ and $m$ values through $l_\text{max}=25$.
\item
$\langle \dot{E} \rangle$ is calculated using Eqs. \eqref{eq:Edot}, \eqref{eq:EMflux}, and \eqref{eq:Gflux}.
\item
Steps (2)-(8) are repeated for a grid of $r_p$ values up to a maximum of $20M$ with grid spacing $\Delta r_p=0.1 M$ ($\sim$150 total grid points).
\item
The \emph{Mathematica} function \emph{Interpolation} (with interpolation order = 11) is used to interpolate $\langle \dot{E} \rangle$ as a function of $r_p$. Figure \ref{fig:intErrFlux} demonstrates that the interpolation has a maximum relative error of $10^{-7}$.
\item
The \emph{Mathematica} function \emph{NDSolve} (with accuracy goal = 7 digits) is used to solve Eqs. \eqref{eq:inspr} and \eqref{eq:inspphi} numerically for arbitrary mass ratios $\epsilon<1$.
\begin{enumerate}[a.]
\item
The initial value for $r_p(t)$ is chosen to be $r_p(0) = 20M$.
\item
The initial value for $\varphi_p(t)$ is chosen to be  $\varphi_p(0)=0$.
\item
When $r_p(t) \le r_\text{ISCO}$, the integration terminates.
\end{enumerate}\item
Steps (1)-(11) are repeated for a set of $Q/M$ values: $0,10^{-7},10^{-6},10^{-4},0.001,0.01,0.1,0.3,0.5,0.6,0.9$
\end{enumerate}

\section{Results}
\label{sec:results}

\subsection{Inspiral dynamics and electromagnetic radiation}

\begin{figure}
\includegraphics[width=3.35in]{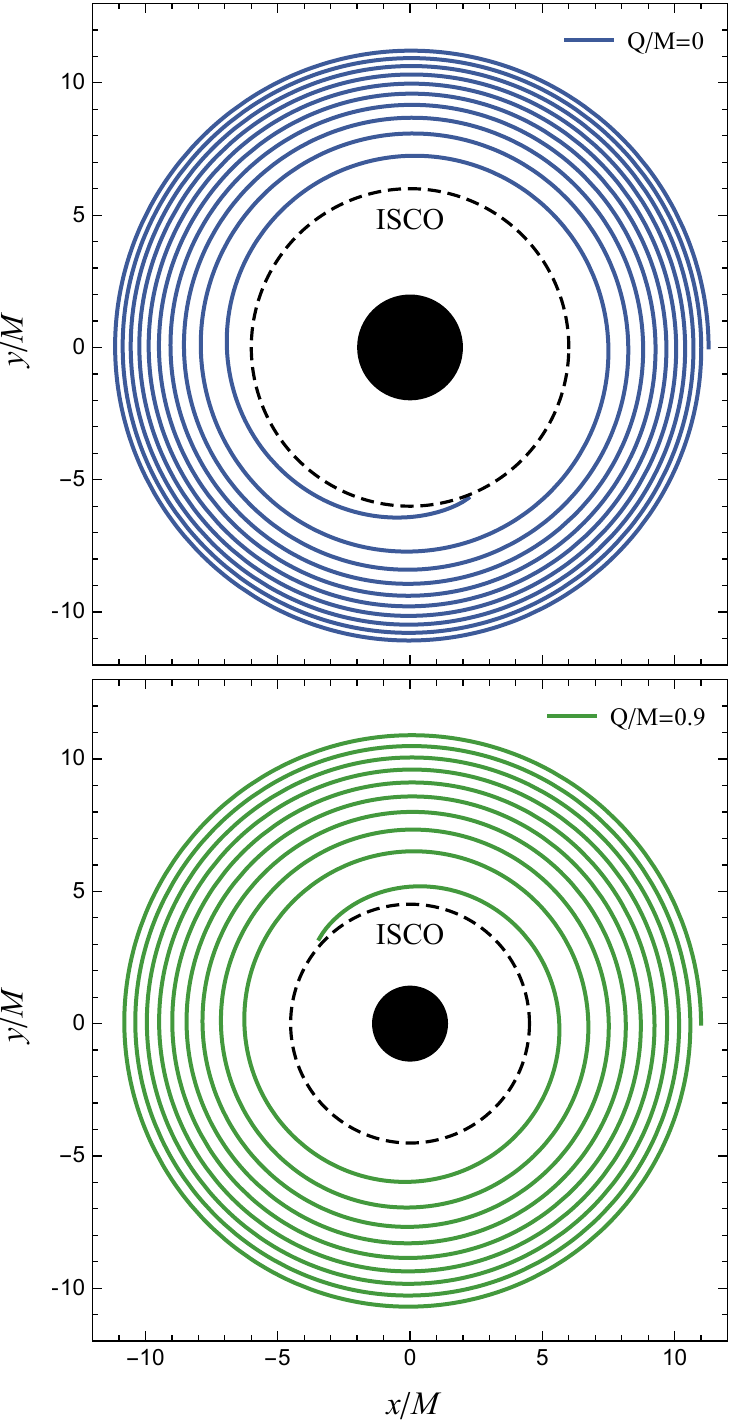}
\caption{\label{fig:insp}Comparison of inspiral trajectories for binaries with two differently charged central black holes. The charges are $Q=0$ and $Q=0.9M$. Both inspirals have mass ratios of $\epsilon=0.1$ and central black holes with mass $M$. Notice that $r_+$ (the radius of the event horizon) and $r_\text{ISCO}$ are smaller when the charge is larger. The initial orbital radii are chosen so that they have the same initial angular speed: $r_p^{Q\ne0}(0) = 11 M$, $r_p^{Q=0}(0) = 11.2841 M$. Note that this larger mass ratio stretches the limits of our perturbative scheme, so this comparison should be interpreted accordingly. Similarly, it would be extraordinary to find an astrophysical black hole with $Q=0.9M$, but such a comparison is an effective illustration of how electric charge affects inspiral dynamics.}
\end{figure}

\begin{figure}
\includegraphics[width=3.35in]{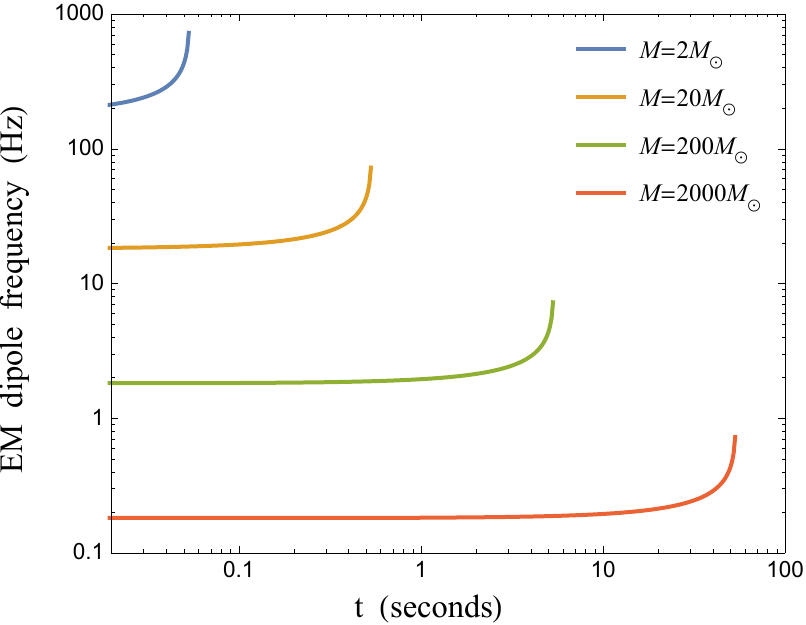}
\caption{\label{fig:EMfreqs}The electromagnetic dipole radiation frequency, approximated by $\nu = \Omega/(2\pi)$, is plotted vs. time for a variety of black hole masses, $M$. To demonstrate that even the highest possible frequencies are too low to be detected by radio observatories, we consider a case that maximizes the angular speed by minimizing the total mass (blue curve). Because the smallest possible masses (therefore the highest possible frequencies) involve $M\sim\mu\sim M_\odot$, we stretch our small mass-ratio approximation beyond its strict regime of applicability as an order of magnitude estimate by choosing $\epsilon=0.5$. Larger masses with even lower frequencies are also shown for comparison. Each depicted inspiral has a charge $Q=0.1M$.}
\end{figure}

Figure \ref{fig:intErrFlux} demonstrates that the electromagnetic energy flux is proportional to $Q^2$ when $Q$ is small. This additional energy flux is one mechanism that could distinguish charged inspiral dynamics. To minimize dephasing during the comparison of gravitational waveforms from differently charged binaries, we require that the two inspirals have the same initial angular speed. Often comparisons are between an inspiral with neutral charge and one with nonzero charge. For these comparisons we hold $\epsilon$ and $M$ constant. According to Eq. \eqref{eq:Omega}, the initial orbital radii must satisfy the following equation to have the same initial angular speed:
\begin{align}
\label{eq:match}
r_p^{Q=0}(0) &= \left(\frac{M r_p^{Q\ne 0}(0)^4}{M r_p^{Q\ne 0}(0) - Q^2}  \right)^{1/3} .
\end{align}
Figure \ref{fig:insp} illustrates how electric charge affects the inspiral dynamics of charged binary systems with matched initial angular speed.

Considering that this system emits light, it is natural to consider possible observations in the electromagnetic spectrum. Ignoring nearby charged particles, charged binary systems radiate electromagnetic waves with extremely low radio frequencies. The frequency is maximized in the case of stellar mass binary components. Existing radio observatories are capable of measuring signals with frequencies as low as a few MHz. Figure \ref{fig:EMfreqs} demonstrates that charged binary systems emit dipole radiation in the kHz range when $M\simeq\mu\simeq M_\odot$ (where our small mass-ratio approximation breaks down). In any other scenario the frequency is lower. Even considering higher harmonics, significant advancements in low frequency radio astronomy would be necessary to detect radio signals from charged binary systems. Potentially small amplitudes are another probable observational challenge. A more likely mechanism of electromagnetic radiation associated with charged binary systems is acceleration of nearby charged particles via the Lorentz force \cite{Zhan16}.

\subsection{Gravitational wave dephasing}

\begin{figure}
\includegraphics[width=3.35in]{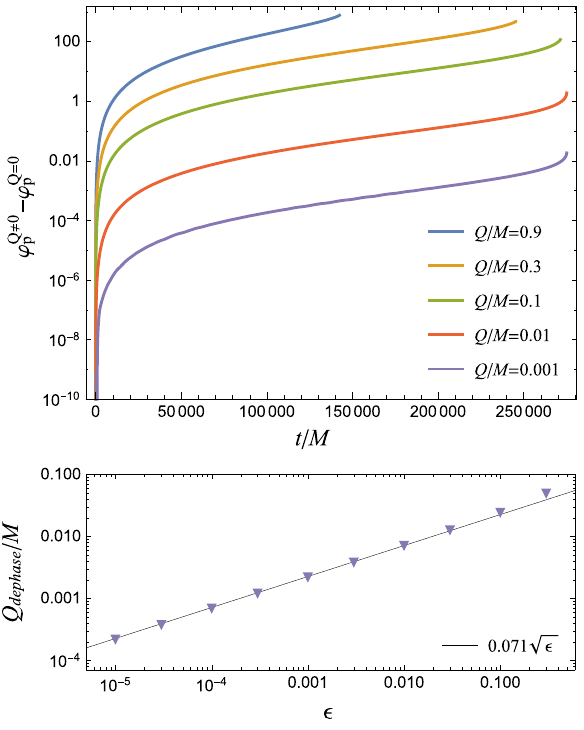}
\caption{\label{fig:deltaPhiDephase} Top: The phase difference between inspirals with and without charge is plotted as a function of time for various $Q/M$ values. The mass ratio of each inspiral is $\epsilon = 0.01$. The total accumulated phase difference, $\Delta \varphi_{\text{total}}$, is approximated by the following equation when $Q$ and $\epsilon$ are small: $\Delta \varphi_{\text{total}} \simeq 198\, \epsilon^{-1} (Q/M)^2$. We observe that the phase of charged inspirals accumulates faster than that of uncharged inspirals. Bottom: The charge threshold above which $\Delta \varphi_\text{total}$ exceeds 1 radian ($Q_\text{dephase}$) is plotted vs. mass ratio ($\epsilon$). We determine $Q_\text{dephase}$ for a given $\epsilon$ by computing a variety of inspirals with different $Q$ values and observing the behavior of $\Delta \varphi_\text{total}$. Note that when $Q$ or $\epsilon$ are large, Eqs. \eqref{eq:totalPhase} and \eqref{eq:dephase} have reduced validity.}
\end{figure}

Gravitational wave detectors use theoretical waveform templates to extract signals from noisy data via matched filtering. Source characteristics are also determined through this process. In order to make successful detections with accurate source parameter estimation, the accumulated waveform phase difference between theoretical models and gravitational wave signals must be less than approximately one radian. For systems studied in this work, the phase of the waveform is calculated from the azimuth of the smaller binary component. Therefore, we estimate the importance of including charge in theoretical waveform templates by calculating the orbital phase difference between charged and uncharged inspirals with matched initial angular speeds. The initial separation could be chosen to reflect a signal's entrance into a detector's passband, but instead we will adopt a uniform initial separation $r_p(0)=20 M$ as an approximation. In actuality we fix $r_p^{Q=0}(0)=20 M$ and determine $r_p^{Q\ne 0}(0)$ from Eq. \eqref{eq:match}.

Figure \ref{fig:deltaPhiDephase} shows the orbital phase difference between charged and uncharged inspirals as a function of time for a given mass ratio. We repeated the analysis of Fig. \ref{fig:deltaPhiDephase} for a wide range of mass ratios. We observe that the phase of charged inspirals accumulates faster than that of uncharged inspirals. We also observe that, when $Q$ and $\epsilon$ are small, the total accumulated phase difference, $\Delta\varphi_\text{total}$, can be determined from a simple empirical relationship
\begin{align}
\label{eq:totalPhase}
\Delta\varphi_\text{total} \simeq 198\, \epsilon^{-1} \left( \frac{Q}{M} \right)^2 ,
\end{align}
where the coefficient is approximate. Observations will be affected when dephasing exceeds $\Delta\varphi_\text{total} \gtrsim 1$. For each $\epsilon$ and $M$ there is a $Q$ for which $\Delta\varphi_\text{total} = 1$ called $Q_\text{dephase}$. We determine the functional form of $Q_\text{dephase}$ by substituting $\Delta\varphi_\text{total} = 1$ into Eq. \eqref{eq:totalPhase}
\begin{align}
\label{eq:dephase}
|Q_\text{dephase}| \simeq 0.071 M \sqrt{\epsilon} .
\end{align}
Figure \ref{fig:deltaPhiDephase} demonstrates that our numerical data approximately satisfy Eq. \eqref{eq:dephase}. As an example consider an extreme mass-ratio inspiral with $M=10^5 \, M_\odot$ and $\mu=1\, M_\odot$ (therefore $\epsilon = 10^{-5}$). For those parameters Eq. \eqref{eq:dephase} predicts $|Q_\text{dephase}| \simeq 3.8 \times 10^{21}$ Coulombs, or $2.4 \times 10^{40}$ electron charges.

\section{Conclusions and future directions}
\label{sec:conc}

In this work we calculated quasicircular inspiral trajectories of a small compact object into a charged black hole for the first time. Our inspiral model incorporates radiation reaction through an adiabatic approximation involving first order in the mass-ratio gravitational and electromagnetic field perturbations. We applied our model to quantify the potential effect of electric charge on gravitational wave observations. We observed that inspirals into a charged black hole evolve more rapidly than comparable inspirals into a neutral black hole. Through analysis of a variety of inspiral configurations, we conclude that charge is an important factor when the amount exceeds the threshold $|Q| \gtrsim 0.071 M \sqrt{\epsilon}$. Above this threshold, waveform templates that exclude charge would noticeably dephase relative to actual signals from charged binaries, and dephasing beyond this level would adversely affect gravitational wave detection and source parameter estimation. We also observe that this system emits light, but at frequencies too low to be detected by modern radio observatories.

A straightforward extension of this project would be to consider the case where both binary components are charged. In that case, the smaller body would be a massive point charge, which would introduce a nonzero current density four-vector, $J^\alpha$. Instead of geodesic motion, the point charge would obey the Lorentz force law. One interesting feature of that extension would be the possibility of opposite or like charges. Another useful extension of this project would be to consider eccentric orbits as small mass-ratio inspirals can be eccentric \cite{Hopm05}.

The analysis of this work largely focuses on waveform templates with total dephasing $\Delta\varphi_\text{total} \simeq 1$ radian over an inspiral. Achieving phase accuracy at that level requires waveform templates to venture beyond the adiabatic approximation. These postadiabatic effects rely on self-force calculations in charged black hole spacetimes. Calculating the self-force would require significant extensions of this work including consideration of $m=0$ perturbations and incorporation of a local regularization scheme. Knowledge of gravitational and electromagnetic self-forces would also facilitate analysis of the cosmic censorship mechanisms that prevent overcharging \cite{SorcWald17} through extension of recent work \cite{ZimmETC13}.

\acknowledgments
We thank Seth Hopper for helpful discussion and comments. R. Z. and T. O. gratefully acknowledge support from the SURE-Oxford program. R. Z. gratefully acknowledges support from the Academic Affairs budget of Oxford College of Emory University, the APS Future of Physics Days Travel Grant, and the APS DGRAV Travel Grant. T. O. gratefully acknowledges support from the Faculty Development Committee of Oxford College of Emory University.

\appendix

\begin{widetext}

\section{Tensor spherical harmonic decomposition}
\label{sec:tsh}

Here we determine the field equations governing the Fourier and spherical harmonic amplitudes of $A^{(1)}_\alpha$ and $g^{(1)}_{\alpha\beta}$ via separation of variables. We perform this analysis in the frequency domain, but the procedures described here can be readily generalized to the time domain. We adopt the conventions of Martel and Poisson \cite{MartPois05} where lowercase Latin indices ($a$,$b$) refer to $t$ and $r$ tensor components and uppercase Latin indices ($A$,$B$) refer to $\theta$ and $\varphi$ tensor components,
\begin{align}
A^{(1)}_b(t,r,\theta,\varphi) &= \sum_{lm} a^{lm}_b(r) \,Y^{lm}(\theta,\varphi) \, e^{-i \omega_m t} ,
\\
A^{(1)}_B(t,r,\theta,\varphi) &= \sum_{lm} \left[a^{lm}_\sharp(r) \, Y_B^{lm}(\theta,\varphi) + a_{lm}^\text{odd}(r) \, X_B^{lm}(\theta,\varphi)\right] e^{-i \omega_m t} ,
\\
g^{(1)}_{ab}(t,r,\theta,\varphi) &= \sum_{lm} h^{lm}_{ab}(r) \,Y^{lm}(\theta,\varphi) \, e^{-i \omega_m t} ,
\\
g^{(1)}_{aB}(t,r,\theta,\varphi) &= \sum_{lm} \left[ j^{lm}_{a}(r) \,Y_B^{lm}(\theta,\varphi) +  h^{lm}_{a}(r) \,X_B^{lm}(\theta,\varphi) \right] e^{-i \omega_m t} ,
\\
g^{(1)}_{AB}(t,r,\theta,\varphi) &= \sum_{lm} \left[ r^2 K^{lm}(r)\, \Omega_{AB}(\theta,\varphi)\, Y^{lm}(\theta,\varphi) + r^2 G^{lm}(r) \, Y^{lm}_{AB}(\theta,\varphi) +  h_2^{lm}(r) \,X_{AB}^{lm}(\theta,\varphi) \right] e^{-i \omega_m t} .
\end{align}
Here $Y^{lm}$ is the spherical harmonic, $Y^{lm}_B$ is the even-parity vector spherical harmonic, $X^{lm}_B$ is the odd-parity vector spherical harmonic, $Y^{lm}_{AB}$ is the even-parity tensor spherical harmonic, $X^{lm}_{AB}$ is the odd-parity tensor spherical harmonic, and $\Omega_{AB}$ is the 2-sphere metric. Because the source is periodic we use a Fourier series to describe the time dependence of $A^{(1)}_\alpha$ and $g^{(1)}_{\alpha\beta}$. For circular motion the source behavior fixes the angular frequency as $\omega_m = m \Omega$. The odd-parity spherical harmonic amplitudes (radial functions) are $a_{lm}^\text{odd}$, $h^{lm}_{t}$, $h^{lm}_{r}$, and $h^{lm}_{2}$, and the even-parity spherical harmonic amplitudes are $a^{lm}_t$, $a^{lm}_r$, $a^{lm}_\sharp$, $h^{lm}_{tt}$, $h^{lm}_{tr}$, $h^{lm}_{rr}$, $j^{lm}_{t}$, $j^{lm}_{r}$, $K^{lm}$, and $G^{lm}$. The metric perturbation is not unique due to the 4 degrees of gauge freedom. We adopt the Regge-Wheeler gauge where $h^{lm}_2=j^{lm}_t=j^{lm}_r=G^{lm}=0$. Similarly, there is 1 degree of electromagnetic gauge freedom, and we adopt the gauge where $a^{lm}_\sharp=0$. 

In terms of these spherical harmonic amplitudes, Maxwell's equations (with $J^\alpha=0$) are
\begin{align}
0 &= f^2\frac{d^2 a_{lm}^\text{odd}}{dr^2} +\frac{2f(Mr-Q^2)}{r^3}\frac{d a_{lm}^\text{odd}}{dr} + \left( \omega_m^2 -\frac{f}{r^2}l(l+1) \right) a_{lm}^\text{odd} - \frac{f Q}{r^2}\frac{d h_t^{lm}}{dr} + \frac{2fQ}{r^3} h_t^{lm} - \frac{i\omega_m f Q}{r^2} h_r^{lm} ,
\label{eq:odd1}
\\
\label{eq:even1}
0 &= -i \omega_m \frac{da_t^{lm}}{dr} + \frac{i\omega_m Q}{2 r^2f} h_{tt}^{lm} - \frac{i\omega_m f Q}{2 r^2} h_{rr}^{lm} + \frac{i\omega_m Q}{r^2} K^{lm} + \left( \omega_m^2-\frac{f}{r^2}l(l+1) \right) a_r^{lm} ,
\\
0 &= f^2 r^3 \frac{da_r^{lm}}{dr} + i\omega_m r^3 a_t^{lm} - 2f(Q^2-Mr) a_r^{lm} ,
\\
0 &= f^2\frac{d^2 a_t^{lm}}{dr^2} +\frac{2f^2}{r}\frac{d a_t^{lm}}{dr} + i\omega_m f^2 \frac{da_r^{lm}}{dr} -\frac{fl(l+1)}{r^2} a_t^{lm} + \frac{2i\omega_m f^2}{r} a_r^{lm}  \notag
\\&\qquad\qquad\qquad\qquad -\frac{fQ}{2r^2}\frac{d h_{tt}^{lm}}{dr} + \frac{f^3Q}{2r^2} \frac{d h_{rr}^{lm}}{dr} - \frac{f^2Q}{r^2} \frac{dK^{lm}}{dr} -\frac{Q(Q^2-M r)}{r^5}  \left( h_{tt}^{lm} + f^2 h_{rr}^{lm} \right) ,
\label{eq:even3}
\end{align}
and the linearized Einstein equations are
\begin{align}
&\mathcal{P}^t_{lm} = \frac{d^2h_t^{lm}}{dr^2} + i\omega_m \frac{dh_r^{lm}}{dr} +\frac{2i\omega_m}{r} h_r^{lm} - \frac{l(l+1)r^2-4 M r+2 Q^2}{f r^4} h_t^{lm} - \frac{4Q}{r^2}\frac{da^\text{odd}_{lm}}{dr} ,
\label{eq:odd2}
\\
&\mathcal{P}^r_{lm} = i\omega_m \frac{dh_t^{lm}}{dr}-\frac{2 i \omega_m}{r} h_t^{lm} - \left(\omega_m^2 -\frac{f}{r^2}(l+2)(l-1) \right) h_r^{lm} - \frac{4i\omega_m Q}{r^2} a^\text{odd}_{lm} ,
\label{eq:odd3}
\\
&\mathcal{P}_{lm} = f\frac{dh_r^{lm}}{dr} - \frac{2(Q^2-M r)}{r^3} h_r^{lm} + \frac{i\omega_m}{f}h_t^{lm},
\label{eq:odd4}
\\
\label{eq:even4}
&\mathcal{Q}^{tt}_{lm} = -\frac{d^2K^{lm}}{dr^2} - \frac{2Q^2+r(3r-5M)}{fr^3}\frac{dK^{lm}}{dr}+\frac{f}{r} \frac{dh_{rr}^{lm}}{dr}+\frac{(l+2)(l-1)}{2fr^2} K^{lm} \notag
\\&\qquad\qquad\qquad\qquad - \frac{4Q^2-r(4M+r(l(l+1)+2))}{2 r^4} h_{rr}^{lm} -\frac{Q^2}{r^4f^2} h_{tt}^{lm} +\frac{2Q}{r^2 f}\frac{da_{t}^{lm}}{dr} + \frac{2i\omega_m Q}{r^2f} a_r^{lm} ,
\\
&\mathcal{Q}^{tr}_{lm} = -i\omega_m \frac{dK^{lm}}{dr}-\frac{l(l+1)}{2r^2} h^{lm}_{tr} +\frac{i\omega_m f}{r} h_{rr}^{lm} -\frac{i\omega_m (2Q^2+r(r-3M))}{r^3f} K^{lm} ,
\\
&\mathcal{Q}^{rr}_{lm} = \frac{f(r-M)}{r^2}\frac{dK^{lm}}{dr} - \frac{f}{r}\frac{dh_{tt}^{lm}}{dr}+\left(\omega_m^2 - \frac{f(l+2)(l-1)}{2r^2}\right)K^{lm} -\frac{f^2}{r^2}h_{rr}^{lm}  \notag
\\&\qquad\qquad\qquad\qquad - \frac{2Q^2-r(4M+rl(l+1))}{2r^4} h_{tt}^{lm} - \frac{2i\omega_m f}{r} h_{tr}^{lm} -\frac{2fQ}{r^2}\frac{da_t^{lm}}{dr} - \frac{2i\omega_m fQ}{r^2}a_r^{lm} ,
\\
&\mathcal{Q}^t_{lm} = -\frac{d h_{tr}^{lm}}{dr}-i \omega_m h_{rr}^{lm} + \frac{2(Q^2-M r)}{r^3f}h_{tr}^{lm} -\frac{i\omega_m}{f} K^{lm} + \frac{4 Q}{r^2}a_r^{lm} ,
\\
&\mathcal{Q}^r_{lm} = \frac{d h_{tt}^{lm}}{dr} -f \frac{dK^{lm}}{dr} - \frac{r-M}{r^2f} h_{tt}^{lm} + i\omega_m h_{tr} +\frac{f(r-M)}{r^2} h_{rr}^{lm} - \frac{4 Q}{r^2} a_t^{lm} ,
\\
&\mathcal{Q}^\flat_{lm} =  f \frac{d^2K^{lm}}{dr^2} - \frac{d^2h_{tt}^{lm}}{dr^2} +\frac{2(r-M)}{r^2}\frac{dK^{lm}}{dr} -\left( \frac{2}{r} -\frac{r-M}{r^2f}\right)\frac{dh_{tt}^{lm}}{dr}-\frac{f(r-M)}{r^2}\frac{dh_{rr}^{lm}}{dr} \notag
\\&\qquad\qquad\qquad\qquad -2i\omega_m \frac{dh_{tr}^{lm}}{dr} +\frac{l(l+1)}{2r^2f}\left( h_{tt}^{lm} -f^2 h_{rr}^{lm} \right)+ \frac{2(r-M)(Q^2-Mr)}{f^2 r^5}\left( h_{tt}^{lm} +f^2 h_{rr}^{lm} \right) \notag
\\&\qquad\qquad\qquad\qquad\qquad\qquad\qquad  +\omega_m^2 h_{rr}^{lm} +\frac{\omega_m^2}{f}K^{lm} -\frac{2i\omega_m(r-M)}{r^2f}h_{tr}^{lm}+\frac{4Q}{r^2}\frac{da_t^{lm}}{dr}+\frac{4i\omega_m Q}{r^2} a_r^{lm} ,
\label{eq:even9}
\\
&\mathcal{Q}^\sharp_{lm} = \frac{1}{f} h_{tt}^{lm} - f h_{rr}^{lm} .
\label{eq:even10}
\end{align}
The source terms, $\mathcal{P}^*_{lm}$ and $\mathcal{Q}^*_{lm}$, arise from spherical harmonic decomposition of the stress-energy tensor. It is important to note that our source terms do not include the electromagnetic part of the stress-energy tensor, only the stress energy of the point mass. In this work the source terms represent a point mass following a circular geodesic. They are related to the definitions of Martel and Poisson \cite{MartPois05} by $ P^*_{lm} = \mathcal{P}^*_{lm}e^{-i\omega_m t}$ (for a point mass following a circular geodesic) and similar for $\mathcal{Q}^*_{lm}$. The source terms are related by four conservation laws that follow from the Einstein-Maxwell equations,
\begin{align}
0 &= i\omega_m \mathcal{P}^t_{lm} - \frac{2}{r} \mathcal{P}^r_{lm}+\frac{(l+2)(l-1)}{r^2} \mathcal{P}_{lm} - \frac{d\mathcal{P}^r_{lm}}{dr} ,
\\
0 &= i\omega_m \mathcal{Q}^{tt}_{lm}+\frac{2(M-r)}{r^2 f} \mathcal{Q}^{tr}_{lm}+\frac{l(l+1)}{2r^2}\mathcal{Q}^{t}_{lm} - \frac{d\mathcal{Q}^{tr}}{dr} ,
\\
0 &= \frac{f(Q^2-Mr)}{r^3} \mathcal{Q}^{tt}_{lm}+i\omega_m \mathcal{Q}^{tr}_{lm}+\left( \frac{r-M}{r^2 f}-\frac{3}{r} \right) \mathcal{Q}^{rr}_{lm}+\frac{l(l+1)}{2r^2} \mathcal{Q}^{r}_{lm} +\frac{f}{r} \mathcal{Q}^\flat_{lm} - \frac{d\mathcal{Q}^{rr}_{lm}}{dr} ,
\\
0 &= i\omega_m \mathcal{Q}^{t}_{lm} - \frac{2}{r} \mathcal{Q}^r_{lm}-\mathcal{Q}^\flat_{lm}+\frac{(l+2)(l-1)}{2r^2} \mathcal{Q}^\sharp_{lm}-\frac{d\mathcal{Q}^r_{lm}}{dr} .
\end{align}

Notice that the even-parity and odd-parity perturbations are not coupled to one another. Our goal is to reduce the field equations for each parity to a system of two coupled second-order differential equations. These master equations will describe the gravitational and electromagnetic master functions from which the gravitational and electromagnetic fields are constructed.

\section{Master function decompositions}
\label{sec:decomp}

Equations \eqref{eq:odd1} and \eqref{eq:odd2}-\eqref{eq:odd4} describe the odd-parity perturbations. The odd-parity gravitational master function, $h_{lm}^\text{odd}$, enters through relationships with $h_t^{lm}$ and $h_r^{lm}$,
\begin{align}
h_t^{lm} &= \frac{rf}{2}\frac{dh^\text{odd}_{lm}}{dr} + \frac{f}{2} h^\text{odd}_{lm} - \frac{f r^2}{(l+2)(l-1)} \mathcal{P}^t_{lm} ,
\\
h_r^{lm} &= -\frac{i\omega_m r}{2f} h^\text{odd}_{lm} + \frac{r^2}{f(l+2)(l-1)} \mathcal{P}^r_{lm} .
\end{align}
The homogeneous version of $h^\text{odd}_{lm}$ is equivalent to the master function ``$\pi_g$'' of Moncrief \cite{Monc74a} up to a constant factor. The electromagnetic master function, $a^\text{odd}_{lm}$, has already appeared in the spherical harmonic decomposition. The homogeneous version of $a^\text{odd}_{lm}$ is equivalent to the master function ``$\pi_f$'' of Moncrief \cite{Monc74a} up to a constant factor. Under these definitions Eqs. \eqref{eq:odd1} and \eqref{eq:odd3} govern the odd-parity electromagnetic and gravitational perturbations,
\begin{align}
& \frac{4 f (Mr-Q^2)}{r^2(l+2)(l-1)} \mathcal{P}^t_{lm} - \frac{2 i\omega_m r}{(l+2)(l-1)} \mathcal{P}^r_{lm} + \frac{2 f^2 r}{(l+2)(l-1)} \frac{d\mathcal{P}^t_{lm}}{dr} = \notag
\\&\qquad\qquad\qquad\qquad\qquad\qquad\qquad\qquad \frac{d^2 h^\text{odd}_{lm}}{dr_*^2} + \left( \omega_m^2 - \frac{f(l(l+1)r^2-6Mr+4Q^2)}{r^4} \right) h^\text{odd}_{lm} - \frac{8 f Q}{r^3}a^\text{odd}_{lm} ,
\label{eq:hodd}
\\
\label{eq:aodd}
& \qquad\qquad\qquad 0 = \frac{d^2 a^\text{odd}_{lm}}{dr_*^2} + \left( \omega_m^2 - \frac{f(l(l+1)r^2+4Q^2)}{r^4} \right) a^\text{odd}_{lm} - \frac{fQ(l+2)(l-1)}{2 r^3} h^\text{odd}_{lm} .
\end{align}
It is straightforward to show that if Eq. \eqref{eq:hodd} is satisfied then Eqs. \eqref{eq:odd2} and \eqref{eq:odd4} will also be satisfied. The odd-parity ODE coefficients and source terms are determined by comparing Eq. \eqref{eq:master} with Eqs. \eqref{eq:hodd} and \eqref{eq:aodd},
\begin{align}
&\alpha_{lm}^\text{odd}(r) = -\frac{f(l(l+1)r^2-6Mr+4Q^2)}{r^4} , \qquad\qquad\qquad\qquad\qquad\qquad\qquad\qquad\qquad\; \beta_{lm}^\text{odd}(r) = - \frac{8 f Q}{r^3} ,
\\
&\sigma_{lm}^\text{odd}(r) = - \frac{f(l(l+1)r^2+4Q^2)}{r^4} , \qquad\qquad\qquad\qquad\qquad\qquad\qquad\qquad\qquad\qquad \gamma_{lm}^\text{odd}(r) = - \frac{fQ(l+2)(l-1)}{2 r^3} ,
\\
&S_{lm}^\text{odd}(r) = \frac{4 f (Mr-Q^2)}{r^2(l+2)(l-1)} \mathcal{P}^t_{lm} - \frac{2 i\omega_m r}{(l+2)(l-1)} \mathcal{P}^r_{lm} + \frac{2 f^2 r}{(l+2)(l-1)} \frac{d\mathcal{P}^t_{lm}}{dr}, \qquad\qquad\;\;\;\, Z_{lm}^\text{odd}(r) = 0 .
\end{align}
Notice that $\alpha_{lm}^\text{odd}$ reduces to the Regge-Wheeler potential when $Q=0$. The Dirac delta function coefficients are determined by analyzing $S_{lm}^\text{odd}$ and $Z_{lm}^\text{odd}$ in the case of a point mass following a circular geodesic,
\begin{align}
&B_{lm}^\text{odd} = \frac{32\pi \mu f_p\mathcal{L}(Q^2-r_p^2)}{r_p^4(l+2)(l+1)l(l-1)}\sqrt{(l+m+1)(l-m)}\, Y^{l,m+1}\left(\frac{\pi}{2},0\right)\,, \qquad\qquad\qquad\qquad D_{lm}^\text{odd} = 0 , \qquad
\\
&F_{lm}^\text{odd} = \frac{32\pi \mu f_p^2 \mathcal{L}}{r_p(l+2)(l+1)l(l-1)}\sqrt{(l+m+1)(l-m)}\, Y^{l,m+1}\left(\frac{\pi}{2},0\right)\,, \qquad\qquad\qquad\qquad\, H_{lm}^\text{odd} = 0. \qquad
\end{align}

Equations \eqref{eq:even1}-\eqref{eq:even3} and \eqref{eq:even4}-\eqref{eq:even10} describe the even-parity perturbations. The spherical harmonic amplitudes $h^{lm}_{tt}$, $h^{lm}_{tr}$, and $h^{lm}_{rr}$ are expressed in terms of $K^{lm}$ by forming linear combinations of Eqs. \eqref{eq:even4}--\eqref{eq:even10} and their $r$ derivatives,
\begin{align}
&h_{rr}^{lm} = \frac{r^2}{2f}\frac{d^2K^{lm}}{dr^2}+\frac{r-M}{f^2}\frac{dK^{lm}}{dr}+\left( \frac{r^2 \omega_m^2}{2f^3} -\frac{(l+2)(l-1)}{2f^2} \right)K^{lm} -\frac{2Q}{f^2}\frac{da_t^{lm}}{dr} -\frac{4Q}{rf^2}a_t^{lm} \notag
\\&\qquad\qquad - \frac{2i\omega_m Q}{f^2} a_r^{lm}+\frac{r}{2f}\frac{d\mathcal{Q}^\sharp_{lm}}{dr}+\frac{r^2}{2f} \mathcal{Q}^{tt}_{lm} - \frac{4Q^2-r(12M+r(l(l+1)-4))}{4r^2f^2} \mathcal{Q}^\sharp_{lm} -\frac{r^2}{2f^3}\mathcal{Q}^{rr}_{lm} - \frac{r}{f^2} \mathcal{Q}^r_{lm} ,
\\
&h_{tr}^{lm} = \frac{2}{l(l+1)}\left( i\omega_m r f h_{rr}^{lm} - i\omega_m r^2\frac{dK^{lm}}{dr} - \frac{i\omega_m(2Q^2+r(r-3M))}{r f} K^{lm} -r^2 Q^{tr}_{lm} \right) ,
\\
&h_{tt}^{lm} = f^2 h_{rr}^{lm} +f Q^\sharp_{lm} .
\end{align}
The even-parity gravitational master function, $h_{lm}^\text{even}$, enters through a relationship with $K^{lm}$,
\begin{align}
\label{eq:heven}
&K^{lm} = \left[ 2 r \left(-2 Q^2 r (2 M+\lambda  r)+(\lambda +1) r^3 (3 M+\lambda  r)+2 Q^4\right) \right]^{-1} \bigg( 2 f (\lambda +1) r^3 \left(r (3 M+\lambda  r)-2 Q^2\right) \frac{dh^\text{even}_{lm}}{dr} \notag
\\&\qquad\qquad\qquad -(\lambda + 1) \left[-2 r^2 \left(6 M^2+3 \lambda  M r+\lambda  (\lambda +1) r^2\right)+2 Q^2 r (11 M+2 (\lambda -1) r)-8 Q^4\right] h^\text{even}_{lm} \notag
\\&\qquad\qquad\qquad\qquad\qquad\qquad +4 r^5 f Q \frac{da_t^{lm}}{dr} + 4 i \omega_m r^5 f Q a_r^{lm} -2 r^7 f^2 \mathcal{Q}^{tt}_{lm} -2 r^3 f Q^2 \mathcal{Q}^\sharp_{lm} \bigg) ,
\end{align}
where $\lambda\equiv (l+2)(l-1)/2$. The homogeneous version of $h^\text{even}_{lm}$ is equivalent to the master function ``$Q$'' of Moncrief \cite{Monc74b} up to a constant factor. Similarly, the spherical harmonic amplitudes $a_t^{lm}$ and $a_r^{lm}$ are expressed in terms of the electromagnetic master function $a^\text{even}_{lm}$,
\begin{align}
\label{eq:aeven}
& a_t^{lm} = -f \frac{d}{dr}\left(a^\text{even}_{lm} +\frac{Q}{2 r} h^\text{even}_{lm} \right) , \qquad\qquad\qquad\qquad\qquad a_r^{lm} = \frac{i \omega_m}{f} \left( a^\text{even}_{lm} +\frac{Q}{2 r} h^\text{even}_{lm} \right) .
\end{align}
The homogeneous version of $a^\text{even}_{lm}$ is equivalent to the master function ``$H$'' of Moncrief \cite{Monc74b} up to a constant factor. It can be shown that, under the definitions of Eqs. \eqref{eq:heven} and \eqref{eq:aeven}, the linearized Einstein-Maxwell equations are satisfied when the following ODE coefficients and source terms are adopted for use with Eq. \eqref{eq:master}:
\begin{align}
&\alpha^\text{even}_{lm}(r) = 2 f \left[ r^4 \left(r (3 M+\lambda  r)-2 Q^2\right)^2 \right]^{-1} \Big(Q^2 r^2 \left(21 M^2+16 \lambda  M r+2 (\lambda -1) \lambda r^2\right) \notag
\\&\qquad\qquad\qquad\qquad\qquad\qquad -r^3 \left(9 M^3+9 \lambda  M^2 r+3 \lambda ^2 M r^2+\lambda ^2 (\lambda +1) r^3\right)-2 Q^4 r (8 M+3 \lambda  r)+4 Q^6 \Big),
\\
&\beta^\text{even}_{lm}(r) = \frac{8 f Q \left(-3 M^2 r+M \left(Q^2+3 r^2\right)+\lambda  (\lambda +2) r^3\right)}{r^2 \left(r (3 M+\lambda  r)-2 Q^2\right)^2} ,
\\
&\sigma^\text{even}_{lm}(r) = -\frac{2 f \left(-2 Q^2 r^2 \left(9 M^2+8 \lambda  M r+(\lambda -1) \lambda  r^2\right)+2 Q^4 r (8 M+3 \lambda  r)+(\lambda +1) r^4 (3 M+\lambda  r)^2-4 Q^6\right)}{r^4 \left(r (3 M+\lambda  r)-2 Q^2\right)^2} ,
\\
&\gamma^\text{even}_{lm}(r) = \frac{ \lambda f Q \left(-3 M^2 r+M \left(Q^2+3 r^2\right)+\lambda  (\lambda +2)
   r^3\right)}{r^2 \left(r (3 M+\lambda  r)-2 Q^2\right)^2} ,
\\
&S^\text{even}_{lm}(r) = \left[ r (3 M+\lambda  r)-2 Q^2 \right]^{-1} \bigg( r^2f\mathcal{Q}^{r}_{lm} + r^3 \mathcal{Q}^{rr}_{lm} - \frac{f(r(3M+\lambda r)-2Q^2)}{r} \mathcal{Q}^\sharp_{lm} - \frac{i\omega_m r^4 f}{\lambda+1} \mathcal{Q}^{tr}_{lm} + \frac{r^4f^3}{\lambda+1}\frac{d\mathcal{Q}^{tt}_{lm}}{dr} \notag
\\&\qquad\qquad\qquad\qquad\qquad\qquad -\frac{f^2 r \left(r^2 \left(12 M^2+3 (\lambda -3) M r+(\lambda -1) \lambda  r^2\right)+2 Q^2 r (5 r-8 M)+4 Q^4\right)}{(\lambda +1) \left(r (3 M+\lambda  r)-2 Q^2\right)} \mathcal{Q}^{tt}_{lm} \bigg) ,
\\
&Z^\text{even}_{lm}(r) =  r Q \left[ 2( r (3 M+\lambda  r)-2 Q^2) \right]^{-1} \bigg( \frac{i \omega_m r^2f }{\lambda +1} \mathcal{Q}^{tr}_{lm} -f \mathcal{Q}^{r}_{lm} - r \mathcal{Q}^{rr}_{lm} - \frac{f^3 r^2 }{\lambda +1} \frac{d \mathcal{Q}^{tt}}{dr} \notag
\\&\qquad\qquad\qquad\qquad\qquad\qquad\qquad\qquad\qquad +\frac{f^2 \left(r \left(2 (\lambda +3) Q^2+\lambda  (\lambda +1) r^2\right)-M \left(2 Q^2+(\lambda +3) r^2\right)\right)}{(\lambda +1) \left(r (3 M+\lambda  r)-2 Q^2\right)} \mathcal{Q}^{tt}_{lm} \bigg) .
\end{align}
Notice that $\alpha^\text{even}_{lm}$ reduces to the Zerilli potential when $Q=0$. The Dirac delta function coefficients are determined by analyzing $S_{lm}^\text{even}$ and $Z_{lm}^\text{even}$ in the case of a point mass following a circular geodesic:
\begin{align}
& B_{lm}^\text{even} = 8 \pi \mu f_p \left[ \mathcal{E} \lambda  (\lambda +1) r_p^3 \left(r_p \left(r_p \lambda + 3 M\right)-2 Q^2\right)^2 \right]^{-1} \Big( \mathcal{L}^2 f_p \left(m^2-\lambda -1\right) \left(r_p \left(3  M+\lambda  r_p\right)-2 Q^2\right)^2 \notag
\\&\qquad\qquad\qquad -  \mathcal{E}^2 \lambda  r_p^2 \left[r_p^2 \left(12 M^2+5 \lambda  M r_p+\lambda  (\lambda +1) r_p^2\right)+Q^2 r_p \left(2 r_p-4 \lambda  r_p-21 M\right)+8 Q^4 \right] \Big) \, Y^{lm}\left(\frac{\pi}{2},0\right)\, ,
\\
& F_{lm}^\text{even} = \frac{8 \pi \mu \,
r_p^2 f_p^2 \mathcal{E}}{(\lambda+1)(r_p(r_p\lambda+3M)-2 Q^2)}\, Y^{lm}\left(\frac{\pi}{2},0\right)\, ,
\\
& D_{lm}^\text{even} = \frac{4 \pi \mu f_p \mathcal{E} Q \left(r_p^2 \left(6 M^2+3 (\lambda +1) M r_p+\lambda  (\lambda +2) r_p^2\right)-Q^2 r_p (14 M+3 \lambda r_p)+6 Q^4\right)}{(\lambda +1) r_p^2 \left(r_p (\lambda r_p+3 M)-2 Q^2\right)^2}\, Y^{lm}\left(\frac{\pi}{2},0\right) \, ,
\\
& H_{lm}^\text{even} = -\frac{4 \pi \mu \, r_p f_p^2 Q \mathcal{E}}{(\lambda+1)(r_p(r_p\lambda+3M)-2 Q^2)}\, Y^{lm}\left(\frac{\pi}{2},0\right)\, .
\end{align}

Past work takes additional steps by defining alternate pairs of master functions from linear combinations of $h_{lm}$ and $a_{lm}$ \cite{Monc74a,Monc74b} (for both even and odd parities). These mixed master function pairs satisfy master equations that are not coupled to one another (which may be simpler mathematically). We choose to sacrifice a level of mathematical simplification in favor of a clearer delineation between gravitational and electromagnetic master functions that instead satisfy a coupled equation, specifically Eq. \eqref{eq:master}.

\section{The dipole ($l=1$) modes}
\label{sec:dipole}

The dipole modes require special treatment. We focus here on the even-parity dipole mode ($l=1$, $m=\pm 1$) because the odd-parity dipole mode ($l=1$, $m=0$) does not radiate for circular orbital motion. The dipole mode requires special treatment because the even-parity tensor spherical harmonic, $Y^{lm}_{AB}$, vanishes when $l=1$. One consequence of vanishing $Y^{1m}_{AB}$ is that Eq. \eqref{eq:even10} does not appear in the system of differential equations. Another consequence of vanishing $Y^{1m}_{AB}$ is that $G^{1m}$ vanishes in all gauges. One of the usual Regge-Wheeler gauge conditions requires that $G^{lm}=0$. The automatic vanishing of $G^{1m}$ relinquishes that degree of gauge freedom. In this work we fix the gauge by instead requiring that $K^{1m}=0$ in addition to $j_t^{1m}=j_r^{1m}=a_\sharp^{1m}=0$. 

The dipole perturbations are described by Eqs. \eqref{eq:even1}-\eqref{eq:even3} and \eqref{eq:even4}-\eqref{eq:even9} with $l=1$. The spherical harmonic amplitudes $h^{1m}_{tt}$ and $h^{1m}_{tr}$ can be expressed in terms of $h^{1m}_{rr}$ by forming linear combinations of Eqs. \eqref{eq:even4}-\eqref{eq:even9},
\begin{align}
h^{1m}_{tt} &= \frac{r^4f}{Q^2-3Mr}\left( \frac{f}{r^3}\left(r^3 \omega_m^2 - M \right) h^{1m}_{rr} -\frac{2Q}{r^2}\frac{d a_t^{1m}}{dr}-\frac{4 Q}{r^3}a_t^{1m}-\frac{2i\omega_m Q}{r^2} a_r^{1m} - \frac{1}{f} \mathcal{Q}^{rr}_{1m}+ i\omega_m r \mathcal{Q}^{tr}_{1m} -\frac{1}{r} \mathcal{Q}^r_{1m} \right) ,
\\
h^{1m}_{tr} &= i\omega_m rf h_{rr}^{1m} - r^2 Q^{tr}_{1m} .
\end{align}
The electromagnetic master function $a^\text{even}_{1m}$ enters through a relationship with $a_t^{1m}$ and $a_r^{1m}$,
\begin{align}
a_t^{1m} &= -f \frac{d a^\text{even}_{1m}}{dr} + \frac{2fQ^2}{r(2Q^2-3Mr)} a^\text{even}_{1m} - \frac{rf^2 Q}{2(2Q^2-3Mr)} h^{1m}_{rr} + \frac{r^3f^2Q}{2(2Q^2-3Mr)} \mathcal{Q}^{tt}_{1m} ,
\\
a_r^{1m} &= \frac{i\omega_m}{f} a^\text{even}_{1m} - \frac{i\omega_m r^2 f Q}{2(2Q^2-3Mr)} h^{1m}_{rr} .
\end{align}
The homogeneous version of $a^\text{even}_{1m}$ is equivalent to the (dipole) master function ``$H$'' of \cite{Monc75} up to a constant factor. Under these definitions, the Einstein-Maxwell equations reduce to the following master equations describing $h^{1m}_{rr}$ and $a^\text{even}_{1m}$:
\begin{align}
\label{eq:hdipole}
&\qquad \frac{r}{f} \mathcal{Q}^{tt}_{1m} = \frac{d h^{1m}_{rr}}{dr} + \left( \frac{3M}{2Q^2-3Mr}+\frac{5r-4M}{r^2f}-\frac{2}{r} \right) h^{1m}_{rr} - \frac{4Q}{r^3f^2} a^\text{even}_{1m} ,
\\
\frac{Q}{2Q^2-3Mr} & \left(\frac{r^3f^3}{2}\frac{d\mathcal{Q}^{tt}_{1m}}{dr} - \frac{rf^2 (2MQ^2-6Q^2 r+3Mr^2)}{2(2Q^2-3Mr)}\mathcal{Q}^{tt}_{1m} - \frac{i\omega_m r^3 f}{2} \mathcal{Q}^{tr}_{1m} + \frac{r^2}{2}\mathcal{Q}^{rr}_{1m} + \frac{rf}{2}\mathcal{Q}^{r}_{1m} \right) \notag
\\&\qquad\qquad\qquad\qquad = \frac{d^2 a_{1m}^\text{even}}{dr_*^2} + \left(\omega_m^2 +\frac{2f(4Q^6-16MQ^4 r+18M^2 Q^2 r^2-9M^2r^4)}{r^4(2Q^2-3Mr)^2} \right) a_{1m}^\text{even} .
\label{eq:adipole}
\end{align}
Notice that Eq. \eqref{eq:adipole} is not coupled to $h^{1m}_{rr}$. Furthermore, and despite very different derivation procedures, the second component of Eq. \eqref{eq:master} for arbitrary even-parity modes reduces to Eq. \eqref{eq:adipole} for ($l=1$, $m=\pm 1$) modes. The decoupling of Eq. \eqref{eq:master} for ($l=1$, $m=\pm 1$) modes is a consequence of the relationship $\gamma^\text{even}_{1m}=0$. Unfortunately, the first component of Eq. \eqref{eq:master} involving $h_{1m}^\text{even}$ is not obviously related to $h^{1m}_{rr}$. Our strategy is to solve Eq. \eqref{eq:master} as usual for ($l=1$, $m=\pm 1$) modes, then disregard $h_{1m}^\text{even}$ (which has no obvious physical meaning), and use $a_{1m}^\text{even}$ to find the electromagnetic dipole energy flux. The dipole metric perturbations are then recovered through Eq. \eqref{eq:hdipole},
\begin{align}
\label{eq:hdipole2}
\frac{8\pi \mu \mathcal{E}}{r_p f_p^2} Y^{1m}\left(\frac{\pi}{2},0\right)\, \delta(r-r_p) + \frac{4Q}{r^3f^2} a^\text{even}_{1m} = \frac{d h^{1m}_{rr}}{dr} + \left( \frac{3M}{2Q^2-3Mr}+\frac{5r-4M}{r^2f}-\frac{2}{r} \right) h^{1m}_{rr} ,
\end{align}
where we have specialized to a circular geodesic and we now treat $a^\text{even}_{1m}$ as a source term. Equation \eqref{eq:hdipole2} has one homogeneous solution, $h^{1m}_{rr,H}$,
\begin{align}
h^{1m}_{rr,H} &= \frac{2Q^2-3 M r}{r^3 f^{5/2}} \left(\frac{r-r_-}{r-r_+}  \right)^{M/(r_+ - r_-)} .
\end{align}
Notice that $h^{1m}_{rr,H}$ is regular at $r=r_+$, but not at $r=\infty$. The inhomogeneous solution of Eq. \eqref{eq:hdipole2} can be expressed as a piecewise function,
\begin{align}
h^{1m}_{rr} &= h^{1m}_{rr,I+} \Theta(r-r_p) + \left( h^{1m}_{rr,I-} + C_{1m}^{rr,H}  h^{1m}_{rr,H}\right) \Theta(r_p-r) .
\end{align}
The function $h^{1m}_{rr,I+}$ represents an inhomogeneous solution that is valid when $r>r_p$, while the function $h^{1m}_{rr,I-}$ represents an inhomogeneous solution that is valid when $r<r_p$,
\begin{align}
h^{1m}_{rr,I\pm} &= h^{1m}_{rr,H} \int_{\pm\infty}^{r_*} \frac{4 Q C_{1m}^{1\pm}a_{1m}^{1\pm}}{r^3 f^2 h^{1m}_{rr,H}} f dr_* ,
\end{align}
where the integrand can be considered a function of $r_*$. The coefficient $C^{1m}_{rr,H}$ is fixed to satisfy the jump condition implied by the Dirac delta function source term
\begin{align}
C_{1m}^{rr,H} = -\frac{1}{h^{1m}_{rr,H}\big|_{r=r_p}}\left(\frac{8\pi \mu \mathcal{E}}{r_p f_p^2} Y_1^m\left(\frac{\pi}{2},0\right) + h^{1m}_{rr,I+}\big|_{r=r_p} - h^{1m}_{rr,I-}\big|_{r=r_p} \right) .
\end{align}
The homogeneous solution is restricted to the region $r<r_p$ to ensure regularity at $r=\infty$.

\section{Boundary expansions}
\label{sec:boundary}

The outgoing homogeneous solutions are expanded in the following form when $r\gg |\omega_m|^{-1}$:
\begin{align}
\label{eq:infExp}
\left[ \begin{array}{c} h_{lm}(r_* \rightarrow +\infty) \\ a_{lm}(r_* \rightarrow +\infty) \end{array} \right] \simeq e^{+i\omega_m r_*}\sum_{j=0}^{j_\text{max}} \frac{1}{(\omega_m r)^{j}} \left[ \begin{array}{c} b^{lm}_j \\ c^{lm}_j \end{array} \right] .
\end{align}
The expansion coefficients, $b^{lm}_j$ and $c^{lm}_j$, are found via the method of Frobenius. The starting coefficients in the expansion, $b^{lm}_0$ and $c^{lm}_0$, can be chosen freely. Higher order coefficients are determined by recurrence relations. As an example we present recurrence relations describing the odd-parity expansion coefficients,
\begin{align}
& 2 j \,b^{lm}_j = \big[12 \omega _m M (j-1) -i (j+l) (j-l-1) \big] b^{lm}_{j-1} +8 i \omega _m Q c^{lm}_{j-2} -48 i \omega _m^2 M Q c^{lm}_{j-3} +24 i \omega _m^3 Q \big[4 M^2+Q^2\big] c^{lm}_{j-4} \notag
\\
& -2 \omega _m \big[3 \omega _m (j-2) (4 M^2+Q^2)-i M \left(j (4 j-11)-3 (l^2+l-1)\right)\big] b^{lm}_{j-2} - 32 i \omega _m^4 M Q \big[2 M^2+3 Q^2\big] c^{lm}_{j-5} \notag
\\
& -i \omega _m^2 \big[12 M^2 \big(j (2 j-9)-(l-2) (l+3)\big)+Q^2 \big(2 j (2 j-9)-3 l (l+1)+14\big)+8 i \omega _m M (j-3) (2 M^2+3 Q^2)\big] b^{lm}_{j-3} \notag
\\
& +2 i \omega _m^3 \big[4 M^3 \big(j (4 j-25)-l (l+1)+27 \big)+3 M Q^2 \big(j (4 j-25)-2 l (l+1)+29 \big)+3 i \omega _m Q^2 (j-4) (4 M^2+Q^2) \big] b^{lm}_{j-4} \notag
\\
& -i \omega _m^4 \big[ 12 M^2 Q^2 \big(4 j (j-8)-l (l+1)+50\big)+3 Q^4 \big(2 j (j-8)-l (l+1)+26\big)+12 i \omega _m M Q^4 (j-5)+16 M^4 (j-6) (j-2) \big] b^{lm}_{j-5} \notag
\\
& +24 i \omega _m^5 Q^3 \big[4 M^2+Q^2\big] c^{lm}_{j-6} +2 i \omega _m^5 Q^2 \big[3 M Q^2 \big(j (4 j-39)-l (l+1)+79\big)+i \omega _m Q^4 (j-6)+4 M^3 (j-7) (4 j-11)\big] b^{lm}_{j-6} \notag
\\
& -48 i \omega _m^6 M Q^5 c^{lm}_{j-7} -2 i \omega _m^6 Q^4 \big[Q^2 \big(4 j^2-46 j-l (l+1)+114\big)+12 M^2 (j-8) (2 j-7)\big] b^{lm}_{j-7} \notag
\\
& +8 i \omega _m^7 Q^7 c^{lm}_{j-8} +2 i \omega _m^7 M Q^6 (j-9) (4 j-17) b^{lm}_{j-8} -i \omega _m^8 Q^8 (j-10) (j-5) b^{lm}_{j-9} \, , \label{eq:bRec}
\\
&4 j \, c^{lm}_j = \big[24 \omega _m M (j-1) -2 i (j+l) (j-l-1)\big] c^{lm}_{j-1} +i \omega _m Q (l+2)(l-1) b^{lm}_{j-2} -6 i \omega _m^2 M Q (l+2)(l-1) b^{lm}_{j-3} \notag
\\
& +4 i \omega _m \big[M \big(j (4 j-11)-3 (l+2)(l-1)\big)+3 i \omega _m (j-2) (4 M^2+Q^2)\big] c^{lm}_{j-2} +3 i \omega _m^3 Q (l+2)(l-1) \big[4 M^2+Q^2\big] b^{lm}_{j-4} \notag
\\
& -2 i \omega _m^2 \big[12 M^2 \big(j (2 j-9)-l (l+1)+9\big)+Q^2 \big(2 j (2 j-9)-3 l (l+1)+14\big)+8 i \omega _m M (j-3) (2 M^2+3 Q^2)\big] c^{lm}_{j-3} \notag
\\
& +4 i \omega _m^3 \big[3 M Q^2 \big(j (4 j-25)-2 (l^2+l-16)\big)+4 M^3 \big(j (4 j-25)-l (l+1)+36\big)+3 i \omega _m Q^2 (j-4) (4 M^2+Q^2)\big] c^{lm}_{j-4} \notag
\\
& -2 i \omega _m^4 \big[12 M^2 Q^2 \big(4 j(j-8)-(l+8)(l-7)\big)+3 Q^4 \big(2 j(j-8)-l (l+1)+26\big)+12 i \omega _m M Q^4 (j-5)+16 M^4 (j-5) (j-3)\big] c^{lm}_{j-5} \notag
\\
& +4 i \omega _m^5 Q^2 \big[3 M Q^2 \big(j (4 j-39)-l (l+1)+82\big)+i \omega _m Q^4 (j-6)+4 M^3 \big(j (4 j-39)+86\big)\big] c^{lm}_{j-6} \notag
\\
& -4 i \omega _m^4 M Q (l+2)(l-1) \big[2 M^2+3 Q^2\big] b^{lm}_{j-5} +3 i \omega _m^5 Q^3 (l+2)(l-1) \big[4 M^2+Q^2\big] b^{lm}_{j-6} -6 i \omega _m^6 M Q^5 (l+2)(l-1) b^{lm}_{j-7} \notag
\\
& -2 i \omega _m^6 Q^4 \big[Q^2 \big(4 j^2-46 j-l (l+1)+114\big)+12 M^2 \big(j (2 j-23)+59\big)\big] c^{lm}_{j-7} +i \omega _m^7 Q^7 (l+2)(l-1) b^{lm}_{j-8} \notag
\\
&+4 i \omega _m^7 M Q^6 \big[j (4 j-53)+156\big] c^{lm}_{j-8}-2 i \omega _m^8 Q^8 (j-5) (j-10) c^{lm}_{j-9} \, . \label{eq:cRec}
\end{align}
The even-parity recurrence relations have a similar form, but with increased complexity. After choosing $b^{lm}_0$ and $c^{lm}_0$, Eqs. \eqref{eq:bRec} and \eqref{eq:cRec} determine all higher order coefficients under the condition $b^{lm}_j = c^{lm}_j = 0$ when $j<0$. The downgoing homogeneous solutions are expanded in the following form when $r-r_+\ll M$:
\begin{align}
\label{eq:horExp}
\left[ \begin{array}{c} h_{lm}(r_* \rightarrow -\infty) \\ a_{lm}(r_* \rightarrow -\infty) \end{array} \right] \simeq e^{-i\omega_m r_*}\sum_{j=0}^{j_\text{max}} (r-r_+)^j \left[ \begin{array}{c} k^{lm}_j \\ p^{lm}_j \end{array} \right] .
\end{align}
The expansion coefficients, $k^{lm}_j$ and $p^{lm}_j$, are similarly found via the method of Frobenius. The starting coefficients in the expansion, $k^{lm}_0$ and $p^{lm}_0$, can be chosen freely. Higher order coefficients are determined by recurrence relations, which we omit for brevity. The dipole expansions follow from the same procedure, but are expanded independently from the other modes due to vanishing terms in the recurrence relations.

Our basis of homogeneous solutions is specified by successive independent choices of starting coefficients in the expansions. The outgoing homogeneous solutions of Eq. \eqref{eq:out} imply the following choices of starting coefficients for use with Eq. \eqref{eq:infExp}
\begin{align}
\left[ \begin{array}{c} b^{lm}_0 \\ c^{lm}_0 \end{array} \right] = \left[ \begin{array}{c} 1 \\ 0 \end{array} \right] \;\;\;\; \longrightarrow \;\;\;\; \left[ \begin{array}{c} h_{lm}^{0+} \\ a_{lm}^{0+} \end{array} \right], \qquad\qquad\qquad\qquad\qquad \left[ \begin{array}{c} b^{lm}_0 \\ c^{lm}_0 \end{array} \right] = \left[ \begin{array}{c} 0 \\ 1 \end{array} \right] \;\;\;\; \longrightarrow \;\;\;\; \left[ \begin{array}{c} h_{lm}^{1+} \\ a_{lm}^{1+} \end{array} \right] .
\end{align}
The downgoing homogeneous solutions of Eq. \eqref{eq:down} imply the following choices of starting coefficients for use with Eq. \eqref{eq:horExp}
\begin{align}
\left[ \begin{array}{c} k^{lm}_0 \\ p^{lm}_0 \end{array} \right] = \left[ \begin{array}{c} 1 \\ 0 \end{array} \right] \;\;\;\; \longrightarrow \;\;\;\; \left[ \begin{array}{c} h_{lm}^{0-} \\ a_{lm}^{0-} \end{array} \right], \qquad\qquad\qquad\qquad\qquad \left[ \begin{array}{c} k^{lm}_0 \\ p^{lm}_0 \end{array} \right] = \left[ \begin{array}{c} 0 \\ 1 \end{array} \right] \;\;\;\; \longrightarrow \;\;\;\; \left[ \begin{array}{c} h_{lm}^{1-} \\ a_{lm}^{1-} \end{array} \right] .
\end{align}
Numerical values for $h^{j\pm}_{lm}$, $a^{j\pm}_{lm}$, and their derivatives are provided by the expansions at an appropriate initial position $r_i$. These numerical data serve as initial values for numerical integration to determine a complete independent set of global homogeneous solutions.

\end{widetext}

\bibliography{charged}

\begin{thebibliography}{56}%
\makeatletter
\providecommand \@ifxundefined [1]{%
 \@ifx{#1\undefined}
}%
\providecommand \@ifnum [1]{%
 \ifnum #1\expandafter \@firstoftwo
 \else \expandafter \@secondoftwo
 \fi
}%
\providecommand \@ifx [1]{%
 \ifx #1\expandafter \@firstoftwo
 \else \expandafter \@secondoftwo
 \fi
}%
\providecommand \natexlab [1]{#1}%
\providecommand \enquote  [1]{``#1''}%
\providecommand \bibnamefont  [1]{#1}%
\providecommand \bibfnamefont [1]{#1}%
\providecommand \citenamefont [1]{#1}%
\providecommand \href@noop [0]{\@secondoftwo}%
\providecommand \href [0]{\begingroup \@sanitize@url \@href}%
\providecommand \@href[1]{\@@startlink{#1}\@@href}%
\providecommand \@@href[1]{\endgroup#1\@@endlink}%
\providecommand \@sanitize@url [0]{\catcode `\\12\catcode `\$12\catcode
  `\&12\catcode `\#12\catcode `\^12\catcode `\_12\catcode `\%12\relax}%
\providecommand \@@startlink[1]{}%
\providecommand \@@endlink[0]{}%
\providecommand \url  [0]{\begingroup\@sanitize@url \@url }%
\providecommand \@url [1]{\endgroup\@href {#1}{\urlprefix }}%
\providecommand \urlprefix  [0]{URL }%
\providecommand \Eprint [0]{\href }%
\providecommand \doibase [0]{http://dx.doi.org/}%
\providecommand \selectlanguage [0]{\@gobble}%
\providecommand \bibinfo  [0]{\@secondoftwo}%
\providecommand \bibfield  [0]{\@secondoftwo}%
\providecommand \translation [1]{[#1]}%
\providecommand \BibitemOpen [0]{}%
\providecommand \bibitemStop [0]{}%
\providecommand \bibitemNoStop [0]{.\EOS\space}%
\providecommand \EOS [0]{\spacefactor3000\relax}%
\providecommand \BibitemShut  [1]{\csname bibitem#1\endcsname}%
\let\auto@bib@innerbib\@empty
\bibitem [{\citenamefont {Harry}\ and\ \citenamefont {$\text{LIGO Scientific
  Collaboration}$}(2010)}]{aLIGO}%
  \BibitemOpen
  \bibfield  {author} {\bibinfo {author} {\bibfnamefont {G.~M.}\ \bibnamefont
  {Harry}}\ and\ \bibinfo {author} {\bibnamefont {$\text{LIGO Scientific
  Collaboration}$}},\ }\href {\doibase 10.1088/0264-9381/27/8/084006}
  {\bibfield  {journal} {\bibinfo  {journal} {Classical and Quantum Gravity}\
  }\textbf {\bibinfo {volume} {27}},\ \bibinfo {pages} {084006} (\bibinfo
  {year} {2010})}\BibitemShut {NoStop}%
\bibitem [{\citenamefont {{Acernese}}\ \emph {et~al.}(2015)\citenamefont
  {{Acernese}} \emph {et~al.}}]{aVIRGO}%
  \BibitemOpen
  \bibfield  {author} {\bibinfo {author} {\bibfnamefont {F.}~\bibnamefont
  {{Acernese}}} \emph {et~al.} (\bibinfo {collaboration} {Virgo
  Collaboration}),\ }\href {\doibase 10.1088/0264-9381/32/2/024001} {\bibfield
  {journal} {\bibinfo  {journal} {Classical and Quantum Gravity}\ }\textbf
  {\bibinfo {volume} {32}},\ \bibinfo {eid} {024001} (\bibinfo {year}
  {2015})}\BibitemShut {NoStop}%
\bibitem [{\citenamefont {Abbott}\ \emph
  {et~al.}(2016{\natexlab{a}})\citenamefont {Abbott} \emph {et~al.}}]{LIGO1}%
  \BibitemOpen
  \bibfield  {author} {\bibinfo {author} {\bibfnamefont {B.~P.}\ \bibnamefont
  {Abbott}} \emph {et~al.} (\bibinfo {collaboration} {LIGO Scientific
  Collaboration and Virgo Collaboration}),\ }\href {\doibase
  10.1103/PhysRevLett.116.061102} {\bibfield  {journal} {\bibinfo  {journal}
  {Phys. Rev. Lett.}\ }\textbf {\bibinfo {volume} {116}},\ \bibinfo {pages}
  {061102} (\bibinfo {year} {2016}{\natexlab{a}})}\BibitemShut {NoStop}%
\bibitem [{\citenamefont {Abbott}\ \emph
  {et~al.}(2016{\natexlab{b}})\citenamefont {Abbott} \emph {et~al.}}]{LIGO2}%
  \BibitemOpen
  \bibfield  {author} {\bibinfo {author} {\bibfnamefont {B.~P.}\ \bibnamefont
  {Abbott}} \emph {et~al.} (\bibinfo {collaboration} {LIGO Scientific
  Collaboration and Virgo Collaboration}),\ }\href {\doibase
  10.1103/PhysRevLett.116.241103} {\bibfield  {journal} {\bibinfo  {journal}
  {Phys. Rev. Lett.}\ }\textbf {\bibinfo {volume} {116}},\ \bibinfo {pages}
  {241103} (\bibinfo {year} {2016}{\natexlab{b}})}\BibitemShut {NoStop}%
\bibitem [{\citenamefont {Abbott}\ \emph
  {et~al.}(2017{\natexlab{a}})\citenamefont {Abbott} \emph {et~al.}}]{LIGO3}%
  \BibitemOpen
  \bibfield  {author} {\bibinfo {author} {\bibfnamefont {B.~P.}\ \bibnamefont
  {Abbott}} \emph {et~al.} (\bibinfo {collaboration} {LIGO Scientific
  Collaboration and Virgo Collaboration}),\ }\href {\doibase
  10.1103/PhysRevLett.118.221101} {\bibfield  {journal} {\bibinfo  {journal}
  {Phys. Rev. Lett.}\ }\textbf {\bibinfo {volume} {118}},\ \bibinfo {pages}
  {221101} (\bibinfo {year} {2017}{\natexlab{a}})}\BibitemShut {NoStop}%
\bibitem [{\citenamefont {Abbott}\ \emph
  {et~al.}(2017{\natexlab{b}})\citenamefont {Abbott} \emph {et~al.}}]{LIGO4}%
  \BibitemOpen
  \bibfield  {author} {\bibinfo {author} {\bibfnamefont {B.~P.}\ \bibnamefont
  {Abbott}} \emph {et~al.} (\bibinfo {collaboration} {LIGO Scientific
  Collaboration and Virgo Collaboration}),\ }\href {\doibase
  10.3847/2041-8213/aa9f0c} {\bibfield  {journal} {\bibinfo  {journal}
  {Astrophysical Journal Letters}\ }\textbf {\bibinfo {volume} {851}},\
  \bibinfo {pages} {L35} (\bibinfo {year} {2017}{\natexlab{b}})}\BibitemShut
  {NoStop}%
\bibitem [{\citenamefont {Abbott}\ \emph
  {et~al.}(2017{\natexlab{c}})\citenamefont {Abbott} \emph {et~al.}}]{LIGO5}%
  \BibitemOpen
  \bibfield  {author} {\bibinfo {author} {\bibfnamefont {B.~P.}\ \bibnamefont
  {Abbott}} \emph {et~al.} (\bibinfo {collaboration} {LIGO Scientific
  Collaboration and Virgo Collaboration}),\ }\href {\doibase
  10.1103/PhysRevLett.119.141101} {\bibfield  {journal} {\bibinfo  {journal}
  {Phys. Rev. Lett.}\ }\textbf {\bibinfo {volume} {119}},\ \bibinfo {pages}
  {141101} (\bibinfo {year} {2017}{\natexlab{c}})}\BibitemShut {NoStop}%
\bibitem [{\citenamefont {Abbott}\ \emph
  {et~al.}(2017{\natexlab{d}})\citenamefont {Abbott} \emph {et~al.}}]{LIGO6}%
  \BibitemOpen
  \bibfield  {author} {\bibinfo {author} {\bibfnamefont {B.~P.}\ \bibnamefont
  {Abbott}} \emph {et~al.} (\bibinfo {collaboration} {LIGO Scientific
  Collaboration and Virgo Collaboration}),\ }\href {\doibase
  10.1103/PhysRevLett.119.161101} {\bibfield  {journal} {\bibinfo  {journal}
  {Phys. Rev. Lett.}\ }\textbf {\bibinfo {volume} {119}},\ \bibinfo {pages}
  {161101} (\bibinfo {year} {2017}{\natexlab{d}})}\BibitemShut {NoStop}%
\bibitem [{\citenamefont {Abbott}\ \emph
  {et~al.}(2016{\natexlab{c}})\citenamefont {Abbott} \emph
  {et~al.}}]{LIGOgrTest}%
  \BibitemOpen
  \bibfield  {author} {\bibinfo {author} {\bibfnamefont {B.~P.}\ \bibnamefont
  {Abbott}} \emph {et~al.} (\bibinfo {collaboration} {LIGO Scientific
  Collaboration and Virgo Collaboration}),\ }\href {\doibase
  10.1103/PhysRevLett.116.221101} {\bibfield  {journal} {\bibinfo  {journal}
  {Phys. Rev. Lett.}\ }\textbf {\bibinfo {volume} {116}},\ \bibinfo {pages}
  {221101} (\bibinfo {year} {2016}{\natexlab{c}})}\BibitemShut {NoStop}%
\bibitem [{\citenamefont {Abbott}\ \emph
  {et~al.}(2016{\natexlab{d}})\citenamefont {Abbott} \emph {et~al.}}]{LIGOapj}%
  \BibitemOpen
  \bibfield  {author} {\bibinfo {author} {\bibfnamefont {B.~P.}\ \bibnamefont
  {Abbott}} \emph {et~al.} (\bibinfo {collaboration} {LIGO Scientific
  Collaboration and Virgo Collaboration}),\ }\href {\doibase
  10.3847/2041-8205/818/2/L22} {\bibfield  {journal} {\bibinfo  {journal}
  {Astrophysical Journal Letters}\ }\textbf {\bibinfo {volume} {818}},\
  \bibinfo {pages} {L22} (\bibinfo {year} {2016}{\natexlab{d}})}\BibitemShut
  {NoStop}%
\bibitem [{\citenamefont {Abbott}\ \emph
  {et~al.}(2017{\natexlab{e}})\citenamefont {Abbott} \emph {et~al.}}]{LIGOns}%
  \BibitemOpen
  \bibfield  {author} {\bibinfo {author} {\bibfnamefont {B.~P.}\ \bibnamefont
  {Abbott}} \emph {et~al.} (\bibinfo {collaboration} {LIGO Scientific
  Collaboration and Virgo Collaboration}),\ }\href {\doibase
  10.3847/2041-8213/aa93fc} {\bibfield  {journal} {\bibinfo  {journal}
  {Astrophysical Journal Letters}\ }\textbf {\bibinfo {volume} {850}},\
  \bibinfo {pages} {L40} (\bibinfo {year} {2017}{\natexlab{e}})}\BibitemShut
  {NoStop}%
\bibitem [{\citenamefont {Abbott}\ \emph
  {et~al.}(2016{\natexlab{e}})\citenamefont {Abbott} \emph
  {et~al.}}]{LIGOsensitive}%
  \BibitemOpen
  \bibfield  {author} {\bibinfo {author} {\bibfnamefont {B.~P.}\ \bibnamefont
  {Abbott}} \emph {et~al.} (\bibinfo {collaboration} {LIGO Scientific
  Collaboration and Virgo Collaboration}),\ }\href {\doibase
  10.1007/lrr-2016-1} {\bibfield  {journal} {\bibinfo  {journal} {LRR}\
  }\textbf {\bibinfo {volume} {19}},\ \bibinfo {pages} {1} (\bibinfo {year}
  {2016}{\natexlab{e}})}\BibitemShut {NoStop}%
\bibitem [{\citenamefont {Aso}\ \emph {et~al.}(2013)\citenamefont {Aso},
  \citenamefont {Michimura}, \citenamefont {Somiya}, \citenamefont {Ando},
  \citenamefont {Miyakawa}, \citenamefont {Sekiguchi}, \citenamefont
  {Tatsumi},\ and\ \citenamefont {Yamamoto}}]{KAGRA}%
  \BibitemOpen
  \bibfield  {author} {\bibinfo {author} {\bibfnamefont {Y.}~\bibnamefont
  {Aso}}, \bibinfo {author} {\bibfnamefont {Y.}~\bibnamefont {Michimura}},
  \bibinfo {author} {\bibfnamefont {K.}~\bibnamefont {Somiya}}, \bibinfo
  {author} {\bibfnamefont {M.}~\bibnamefont {Ando}}, \bibinfo {author}
  {\bibfnamefont {O.}~\bibnamefont {Miyakawa}}, \bibinfo {author}
  {\bibfnamefont {T.}~\bibnamefont {Sekiguchi}}, \bibinfo {author}
  {\bibfnamefont {D.}~\bibnamefont {Tatsumi}}, \ and\ \bibinfo {author}
  {\bibfnamefont {H.}~\bibnamefont {Yamamoto}} (\bibinfo {collaboration} {The
  KAGRA Collaboration}),\ }\href {\doibase 10.1103/PhysRevD.88.043007}
  {\bibfield  {journal} {\bibinfo  {journal} {Phys. Rev. D}\ }\textbf {\bibinfo
  {volume} {88}},\ \bibinfo {pages} {043007} (\bibinfo {year}
  {2013})}\BibitemShut {NoStop}%
\bibitem [{\citenamefont {Unnikrishnan}(2013)}]{indigo}%
  \BibitemOpen
  \bibfield  {author} {\bibinfo {author} {\bibfnamefont {C.~S.}\ \bibnamefont
  {Unnikrishnan}},\ }\href {\doibase 10.1142/S0218271813410101} {\bibfield
  {journal} {\bibinfo  {journal} {International Journal of Modern Physics D}\
  }\textbf {\bibinfo {volume} {22}},\ \bibinfo {pages} {1341010} (\bibinfo
  {year} {2013})}\BibitemShut {NoStop}%
\bibitem [{\citenamefont {{Martynov}}\ \emph {et~al.}(2016)\citenamefont
  {{Martynov}}, \citenamefont {{Hall}}, \citenamefont {{Abbott}}, \citenamefont
  {{Abbott}}, \citenamefont {{Abbott}}, \citenamefont {{Adams}}, \citenamefont
  {{Adhikari}}, \citenamefont {{Anderson}}, \citenamefont {{Anderson}},
  \citenamefont {{Arai}},\ and\ \citenamefont {et~al.}}]{LIGOfreq}%
  \BibitemOpen
  \bibfield  {author} {\bibinfo {author} {\bibfnamefont {D.~V.}\ \bibnamefont
  {{Martynov}}}, \bibinfo {author} {\bibfnamefont {E.~D.}\ \bibnamefont
  {{Hall}}}, \bibinfo {author} {\bibfnamefont {B.~P.}\ \bibnamefont
  {{Abbott}}}, \bibinfo {author} {\bibfnamefont {R.}~\bibnamefont {{Abbott}}},
  \bibinfo {author} {\bibfnamefont {T.~D.}\ \bibnamefont {{Abbott}}}, \bibinfo
  {author} {\bibfnamefont {C.}~\bibnamefont {{Adams}}}, \bibinfo {author}
  {\bibfnamefont {R.~X.}\ \bibnamefont {{Adhikari}}}, \bibinfo {author}
  {\bibfnamefont {R.~A.}\ \bibnamefont {{Anderson}}}, \bibinfo {author}
  {\bibfnamefont {S.~B.}\ \bibnamefont {{Anderson}}}, \bibinfo {author}
  {\bibfnamefont {K.}~\bibnamefont {{Arai}}}, \ and\ \bibinfo {author}
  {\bibnamefont {et~al.}},\ }\href {\doibase 10.1103/PhysRevD.93.112004}
  {\bibfield  {journal} {\bibinfo  {journal} {Phys. Rev. D}\ }\textbf {\bibinfo
  {volume} {93}},\ \bibinfo {eid} {112004} (\bibinfo {year}
  {2016})}\BibitemShut {NoStop}%
\bibitem [{\citenamefont {{Amaro-Seoane}}\ \emph {et~al.}(2017)\citenamefont
  {{Amaro-Seoane}} \emph {et~al.}}]{LISA}%
  \BibitemOpen
  \bibfield  {author} {\bibinfo {author} {\bibfnamefont {P.}~\bibnamefont
  {{Amaro-Seoane}}} \emph {et~al.},\ }\href@noop {} {\bibfield  {journal}
  {\bibinfo  {journal} {ArXiv e-prints}\ } (\bibinfo {year} {2017})},\ \Eprint
  {http://arxiv.org/abs/1702.00786} {arXiv:1702.00786 [astro-ph.IM]}
  \BibitemShut {NoStop}%
\bibitem [{\citenamefont {Klein}\ \emph {et~al.}(2016)\citenamefont {Klein},
  \citenamefont {Barausse}, \citenamefont {Sesana}, \citenamefont {Petiteau},
  \citenamefont {Berti}, \citenamefont {Babak}, \citenamefont {Gair},
  \citenamefont {Aoudia}, \citenamefont {Hinder}, \citenamefont {Ohme},\ and\
  \citenamefont {Wardell}}]{LISAfreq}%
  \BibitemOpen
  \bibfield  {author} {\bibinfo {author} {\bibfnamefont {A.}~\bibnamefont
  {Klein}}, \bibinfo {author} {\bibfnamefont {E.}~\bibnamefont {Barausse}},
  \bibinfo {author} {\bibfnamefont {A.}~\bibnamefont {Sesana}}, \bibinfo
  {author} {\bibfnamefont {A.}~\bibnamefont {Petiteau}}, \bibinfo {author}
  {\bibfnamefont {E.}~\bibnamefont {Berti}}, \bibinfo {author} {\bibfnamefont
  {S.}~\bibnamefont {Babak}}, \bibinfo {author} {\bibfnamefont
  {J.}~\bibnamefont {Gair}}, \bibinfo {author} {\bibfnamefont {S.}~\bibnamefont
  {Aoudia}}, \bibinfo {author} {\bibfnamefont {I.}~\bibnamefont {Hinder}},
  \bibinfo {author} {\bibfnamefont {F.}~\bibnamefont {Ohme}}, \ and\ \bibinfo
  {author} {\bibfnamefont {B.}~\bibnamefont {Wardell}},\ }\href {\doibase
  10.1103/PhysRevD.93.024003} {\bibfield  {journal} {\bibinfo  {journal} {Phys.
  Rev. D}\ }\textbf {\bibinfo {volume} {93}},\ \bibinfo {pages} {024003}
  (\bibinfo {year} {2016})}\BibitemShut {NoStop}%
\bibitem [{\citenamefont {Manchester}\ and\ \citenamefont
  {$\text{IPTA}$}(2013)}]{PTA}%
  \BibitemOpen
  \bibfield  {author} {\bibinfo {author} {\bibfnamefont {R.~N.}\ \bibnamefont
  {Manchester}}\ and\ \bibinfo {author} {\bibnamefont {$\text{IPTA}$}},\ }\href
  {\doibase 10.1088/0264-9381/30/22/224010} {\bibfield  {journal} {\bibinfo
  {journal} {Classical and Quantum Gravity}\ }\textbf {\bibinfo {volume}
  {30}},\ \bibinfo {pages} {224010} (\bibinfo {year} {2013})}\BibitemShut
  {NoStop}%
\bibitem [{\citenamefont {Hobbs}\ \emph {et~al.}(2010)\citenamefont {Hobbs}
  \emph {et~al.}}]{IPTA}%
  \BibitemOpen
  \bibfield  {author} {\bibinfo {author} {\bibfnamefont {G.}~\bibnamefont
  {Hobbs}} \emph {et~al.},\ }\href {\doibase 10.1088/0264-9381/27/8/084013}
  {\bibfield  {journal} {\bibinfo  {journal} {Classical and Quantum Gravity}\
  }\textbf {\bibinfo {volume} {27}},\ \bibinfo {pages} {084013} (\bibinfo
  {year} {2010})}\BibitemShut {NoStop}%
\bibitem [{\citenamefont {{Barack}}\ and\ \citenamefont
  {{Cutler}}(2004)}]{BaraCutl04}%
  \BibitemOpen
  \bibfield  {author} {\bibinfo {author} {\bibfnamefont {L.}~\bibnamefont
  {{Barack}}}\ and\ \bibinfo {author} {\bibfnamefont {C.}~\bibnamefont
  {{Cutler}}},\ }\href {\doibase 10.1103/PhysRevD.69.082005} {\bibfield
  {journal} {\bibinfo  {journal} {Phys. Rev. D}\ }\textbf {\bibinfo {volume}
  {69}},\ \bibinfo {eid} {082005} (\bibinfo {year} {2004})}\BibitemShut
  {NoStop}%
\bibitem [{\citenamefont {Narayan}(2005)}]{Nara05}%
  \BibitemOpen
  \bibfield  {author} {\bibinfo {author} {\bibfnamefont {R.}~\bibnamefont
  {Narayan}},\ }\href {\doibase 10.1088/1367-2630/7/1/199} {\bibfield
  {journal} {\bibinfo  {journal} {New Journal of Physics}\ }\textbf {\bibinfo
  {volume} {7}},\ \bibinfo {pages} {199} (\bibinfo {year} {2005})}\BibitemShut
  {NoStop}%
\bibitem [{\citenamefont {Wald}(1974)}]{Wald74}%
  \BibitemOpen
  \bibfield  {author} {\bibinfo {author} {\bibfnamefont {R.~M.}\ \bibnamefont
  {Wald}},\ }\href {\doibase 10.1103/PhysRevD.10.1680} {\bibfield  {journal}
  {\bibinfo  {journal} {Phys. Rev. D}\ }\textbf {\bibinfo {volume} {10}},\
  \bibinfo {pages} {1680} (\bibinfo {year} {1974})}\BibitemShut {NoStop}%
\bibitem [{\citenamefont {{Cohen}}\ \emph {et~al.}(1975)\citenamefont
  {{Cohen}}, \citenamefont {{Kegeles}},\ and\ \citenamefont
  {{Rosenblum}}}]{Cohe75}%
  \BibitemOpen
  \bibfield  {author} {\bibinfo {author} {\bibfnamefont {J.~M.}\ \bibnamefont
  {{Cohen}}}, \bibinfo {author} {\bibfnamefont {L.~S.}\ \bibnamefont
  {{Kegeles}}}, \ and\ \bibinfo {author} {\bibfnamefont {A.}~\bibnamefont
  {{Rosenblum}}},\ }\href {\doibase 10.1086/153944} {\bibfield  {journal}
  {\bibinfo  {journal} {Astrophysical Journal}\ }\textbf {\bibinfo {volume}
  {201}},\ \bibinfo {pages} {783} (\bibinfo {year} {1975})}\BibitemShut
  {NoStop}%
\bibitem [{\citenamefont {Lee}\ \emph {et~al.}(2001)\citenamefont {Lee},
  \citenamefont {Lee},\ and\ \citenamefont {van Putten}}]{Lee01}%
  \BibitemOpen
  \bibfield  {author} {\bibinfo {author} {\bibfnamefont {H.~K.}\ \bibnamefont
  {Lee}}, \bibinfo {author} {\bibfnamefont {C.~H.}\ \bibnamefont {Lee}}, \ and\
  \bibinfo {author} {\bibfnamefont {M.~H. P.~M.}\ \bibnamefont {van Putten}},\
  }\href {\doibase 10.1046/j.1365-8711.2001.04401.x} {\bibfield  {journal}
  {\bibinfo  {journal} {Monthly Notices of the Royal Astronomical Society}\
  }\textbf {\bibinfo {volume} {324}},\ \bibinfo {pages} {781} (\bibinfo {year}
  {2001})}\BibitemShut {NoStop}%
\bibitem [{\citenamefont {De~Rujula}\ \emph {et~al.}(1990)\citenamefont
  {De~Rujula}, \citenamefont {Glashow},\ and\ \citenamefont
  {Sarid}}]{RujuETC90}%
  \BibitemOpen
  \bibfield  {author} {\bibinfo {author} {\bibfnamefont {A.}~\bibnamefont
  {De~Rujula}}, \bibinfo {author} {\bibfnamefont {S.}~\bibnamefont {Glashow}},
  \ and\ \bibinfo {author} {\bibfnamefont {U.}~\bibnamefont {Sarid}},\ }\href
  {\doibase 10.1016/0550-3213(90)90227-5} {\bibfield  {journal} {\bibinfo
  {journal} {Nuclear Physics B}\ }\textbf {\bibinfo {volume} {333}},\ \bibinfo
  {pages} {173} (\bibinfo {year} {1990})}\BibitemShut {NoStop}%
\bibitem [{\citenamefont {Sigurdson}\ \emph {et~al.}(2004)\citenamefont
  {Sigurdson}, \citenamefont {Doran}, \citenamefont {Kurylov}, \citenamefont
  {Caldwell},\ and\ \citenamefont {Kamionkowski}}]{SiguETC04}%
  \BibitemOpen
  \bibfield  {author} {\bibinfo {author} {\bibfnamefont {K.}~\bibnamefont
  {Sigurdson}}, \bibinfo {author} {\bibfnamefont {M.}~\bibnamefont {Doran}},
  \bibinfo {author} {\bibfnamefont {A.}~\bibnamefont {Kurylov}}, \bibinfo
  {author} {\bibfnamefont {R.~R.}\ \bibnamefont {Caldwell}}, \ and\ \bibinfo
  {author} {\bibfnamefont {M.}~\bibnamefont {Kamionkowski}},\ }\href {\doibase
  10.1103/PhysRevD.70.083501, 10.1103/PhysRevD.73.089903} {\bibfield  {journal}
  {\bibinfo  {journal} {Phys. Rev. D}\ }\textbf {\bibinfo {volume} {70}},\
  \bibinfo {pages} {083501} (\bibinfo {year} {2004})}\BibitemShut {NoStop}%
\bibitem [{\citenamefont {Cardoso}\ \emph {et~al.}(2016)\citenamefont
  {Cardoso}, \citenamefont {Macedo}, \citenamefont {Pani},\ and\ \citenamefont
  {Ferrari}}]{CardETC16}%
  \BibitemOpen
  \bibfield  {author} {\bibinfo {author} {\bibfnamefont {V.}~\bibnamefont
  {Cardoso}}, \bibinfo {author} {\bibfnamefont {C.~F.~B.}\ \bibnamefont
  {Macedo}}, \bibinfo {author} {\bibfnamefont {P.}~\bibnamefont {Pani}}, \ and\
  \bibinfo {author} {\bibfnamefont {V.}~\bibnamefont {Ferrari}},\ }\href
  {\doibase 10.1088/1475-7516/2016/05/054} {\bibfield  {journal} {\bibinfo
  {journal} {Journal of Cosmology and Astroparticle Physics}\ }\textbf
  {\bibinfo {volume} {1605}},\ \bibinfo {pages} {054} (\bibinfo {year}
  {2016})}\BibitemShut {NoStop}%
\bibitem [{\citenamefont {Zilh{\~a}o}\ \emph {et~al.}(2012)\citenamefont
  {Zilh{\~a}o}, \citenamefont {Cardoso}, \citenamefont {Herdeiro},
  \citenamefont {Lehner},\ and\ \citenamefont {Sperhake}}]{Zilh12}%
  \BibitemOpen
  \bibfield  {author} {\bibinfo {author} {\bibfnamefont {M.}~\bibnamefont
  {Zilh{\~a}o}}, \bibinfo {author} {\bibfnamefont {V.}~\bibnamefont {Cardoso}},
  \bibinfo {author} {\bibfnamefont {C.}~\bibnamefont {Herdeiro}}, \bibinfo
  {author} {\bibfnamefont {L.}~\bibnamefont {Lehner}}, \ and\ \bibinfo {author}
  {\bibfnamefont {U.}~\bibnamefont {Sperhake}},\ }\href {\doibase
  10.1103/PhysRevD.85.124062} {\bibfield  {journal} {\bibinfo  {journal} {Phys.
  Rev. D}\ }\textbf {\bibinfo {volume} {85}},\ \bibinfo {pages} {124062}
  (\bibinfo {year} {2012})}\BibitemShut {NoStop}%
\bibitem [{\citenamefont {Zilh{\~a}o}\ \emph {et~al.}(2014)\citenamefont
  {Zilh{\~a}o}, \citenamefont {Cardoso}, \citenamefont {Herdeiro},
  \citenamefont {Lehner},\ and\ \citenamefont {Sperhake}}]{Zilh14}%
  \BibitemOpen
  \bibfield  {author} {\bibinfo {author} {\bibfnamefont {M.}~\bibnamefont
  {Zilh{\~a}o}}, \bibinfo {author} {\bibfnamefont {V.}~\bibnamefont {Cardoso}},
  \bibinfo {author} {\bibfnamefont {C.}~\bibnamefont {Herdeiro}}, \bibinfo
  {author} {\bibfnamefont {L.}~\bibnamefont {Lehner}}, \ and\ \bibinfo {author}
  {\bibfnamefont {U.}~\bibnamefont {Sperhake}},\ }\href {\doibase
  10.1103/PhysRevD.89.044008} {\bibfield  {journal} {\bibinfo  {journal} {Phys.
  Rev. D}\ }\textbf {\bibinfo {volume} {89}},\ \bibinfo {pages} {044008}
  (\bibinfo {year} {2014})}\BibitemShut {NoStop}%
\bibitem [{\citenamefont {Zhang}(2016)}]{Zhan16}%
  \BibitemOpen
  \bibfield  {author} {\bibinfo {author} {\bibfnamefont {B.}~\bibnamefont
  {Zhang}},\ }\href {\doibase 10.3847/2041-8205/827/2/L31} {\bibfield
  {journal} {\bibinfo  {journal} {Astrophysical Journal Letters}\ }\textbf
  {\bibinfo {volume} {827}},\ \bibinfo {pages} {L31} (\bibinfo {year}
  {2016})}\BibitemShut {NoStop}%
\bibitem [{\citenamefont {Bini}\ \emph {et~al.}(2016)\citenamefont {Bini},
  \citenamefont {Carvalho},\ and\ \citenamefont {Geralico}}]{Bini16}%
  \BibitemOpen
  \bibfield  {author} {\bibinfo {author} {\bibfnamefont {D.}~\bibnamefont
  {Bini}}, \bibinfo {author} {\bibfnamefont {G.~G.}\ \bibnamefont {Carvalho}},
  \ and\ \bibinfo {author} {\bibfnamefont {A.}~\bibnamefont {Geralico}},\
  }\href {\doibase 10.1103/PhysRevD.94.124028} {\bibfield  {journal} {\bibinfo
  {journal} {Phys. Rev. D}\ }\textbf {\bibinfo {volume} {94}},\ \bibinfo
  {pages} {124028} (\bibinfo {year} {2016})}\BibitemShut {NoStop}%
\bibitem [{\citenamefont {{Zerilli}}(1974)}]{Zeri74}%
  \BibitemOpen
  \bibfield  {author} {\bibinfo {author} {\bibfnamefont {F.~J.}\ \bibnamefont
  {{Zerilli}}},\ }\href {\doibase 10.1103/PhysRevD.9.860} {\bibfield  {journal}
  {\bibinfo  {journal} {Phys. Rev. D}\ }\textbf {\bibinfo {volume} {9}},\
  \bibinfo {pages} {860} (\bibinfo {year} {1974})}\BibitemShut {NoStop}%
\bibitem [{\citenamefont {Johnston}\ \emph {et~al.}(1973)\citenamefont
  {Johnston}, \citenamefont {Ruffini},\ and\ \citenamefont {Zerilli}}]{John73}%
  \BibitemOpen
  \bibfield  {author} {\bibinfo {author} {\bibfnamefont {M.}~\bibnamefont
  {Johnston}}, \bibinfo {author} {\bibfnamefont {R.}~\bibnamefont {Ruffini}}, \
  and\ \bibinfo {author} {\bibfnamefont {F.}~\bibnamefont {Zerilli}},\ }\href
  {\doibase 10.1103/PhysRevLett.31.1317} {\bibfield  {journal} {\bibinfo
  {journal} {Phys. Rev. Lett.}\ }\textbf {\bibinfo {volume} {31}},\ \bibinfo
  {pages} {1317} (\bibinfo {year} {1973})}\BibitemShut {NoStop}%
\bibitem [{\citenamefont {Johnston}\ \emph {et~al.}(1974)\citenamefont
  {Johnston}, \citenamefont {Ruffini},\ and\ \citenamefont {Zerilli}}]{John74}%
  \BibitemOpen
  \bibfield  {author} {\bibinfo {author} {\bibfnamefont {M.}~\bibnamefont
  {Johnston}}, \bibinfo {author} {\bibfnamefont {R.}~\bibnamefont {Ruffini}}, \
  and\ \bibinfo {author} {\bibfnamefont {F.}~\bibnamefont {Zerilli}},\ }\href
  {\doibase 10.1016/0370-2693(74)90505-X} {\bibfield  {journal} {\bibinfo
  {journal} {Physics Letters B}\ }\textbf {\bibinfo {volume} {49}},\ \bibinfo
  {pages} {185 } (\bibinfo {year} {1974})}\BibitemShut {NoStop}%
\bibitem [{\citenamefont {Moncrief}(1974{\natexlab{a}})}]{Monc74a}%
  \BibitemOpen
  \bibfield  {author} {\bibinfo {author} {\bibfnamefont {V.}~\bibnamefont
  {Moncrief}},\ }\href {\doibase 10.1103/PhysRevD.9.2707} {\bibfield  {journal}
  {\bibinfo  {journal} {Phys. Rev. D}\ }\textbf {\bibinfo {volume} {9}},\
  \bibinfo {pages} {2707} (\bibinfo {year} {1974}{\natexlab{a}})}\BibitemShut
  {NoStop}%
\bibitem [{\citenamefont {Moncrief}(1974{\natexlab{b}})}]{Monc74b}%
  \BibitemOpen
  \bibfield  {author} {\bibinfo {author} {\bibfnamefont {V.}~\bibnamefont
  {Moncrief}},\ }\href {\doibase 10.1103/PhysRevD.10.1057} {\bibfield
  {journal} {\bibinfo  {journal} {Phys. Rev. D}\ }\textbf {\bibinfo {volume}
  {10}},\ \bibinfo {pages} {1057} (\bibinfo {year}
  {1974}{\natexlab{b}})}\BibitemShut {NoStop}%
\bibitem [{\citenamefont {Moncrief}(1975)}]{Monc75}%
  \BibitemOpen
  \bibfield  {author} {\bibinfo {author} {\bibfnamefont {V.}~\bibnamefont
  {Moncrief}},\ }\href {\doibase 10.1103/PhysRevD.12.1526} {\bibfield
  {journal} {\bibinfo  {journal} {Phys. Rev. D}\ }\textbf {\bibinfo {volume}
  {12}},\ \bibinfo {pages} {1526} (\bibinfo {year} {1975})}\BibitemShut
  {NoStop}%
\bibitem [{\citenamefont {Chandrasekhar}(1979)}]{Chan79}%
  \BibitemOpen
  \bibfield  {author} {\bibinfo {author} {\bibfnamefont {S.}~\bibnamefont
  {Chandrasekhar}},\ }\href {\doibase 10.1098/rspa.1979.0028} {\bibfield
  {journal} {\bibinfo  {journal} {RSPSA}\ }\textbf {\bibinfo {volume} {365}},\
  \bibinfo {pages} {453} (\bibinfo {year} {1979})}\BibitemShut {NoStop}%
\bibitem [{\citenamefont {Gunter}(1980)}]{Gunt80}%
  \BibitemOpen
  \bibfield  {author} {\bibinfo {author} {\bibfnamefont {D.~L.}\ \bibnamefont
  {Gunter}},\ }\href {\doibase 10.1098/rsta.1980.0190} {\bibfield  {journal}
  {\bibinfo  {journal} {Philosophical Transactions of the Royal Society A}\
  }\textbf {\bibinfo {volume} {296}},\ \bibinfo {pages} {497} (\bibinfo {year}
  {1980})}\BibitemShut {NoStop}%
\bibitem [{\citenamefont {Bini}\ \emph {et~al.}(2007)\citenamefont {Bini},
  \citenamefont {Geralico},\ and\ \citenamefont {Ruffini}}]{Bini07}%
  \BibitemOpen
  \bibfield  {author} {\bibinfo {author} {\bibfnamefont {D.}~\bibnamefont
  {Bini}}, \bibinfo {author} {\bibfnamefont {A.}~\bibnamefont {Geralico}}, \
  and\ \bibinfo {author} {\bibfnamefont {R.}~\bibnamefont {Ruffini}},\ }\href
  {\doibase 10.1103/PhysRevD.75.044012} {\bibfield  {journal} {\bibinfo
  {journal} {Phys. Rev. D}\ }\textbf {\bibinfo {volume} {75}},\ \bibinfo
  {pages} {044012} (\bibinfo {year} {2007})}\BibitemShut {NoStop}%
\bibitem [{\citenamefont {Mino}\ \emph {et~al.}(1997)\citenamefont {Mino},
  \citenamefont {Sasaki},\ and\ \citenamefont {Tanaka}}]{MinoSasaTana97}%
  \BibitemOpen
  \bibfield  {author} {\bibinfo {author} {\bibfnamefont {Y.}~\bibnamefont
  {Mino}}, \bibinfo {author} {\bibfnamefont {M.}~\bibnamefont {Sasaki}}, \ and\
  \bibinfo {author} {\bibfnamefont {T.}~\bibnamefont {Tanaka}},\ }\href
  {10.1103/PhysRevD.55.3457} {\bibfield  {journal} {\bibinfo  {journal} {Phys.
  Rev. D}\ }\textbf {\bibinfo {volume} {55}},\ \bibinfo {pages} {3457}
  (\bibinfo {year} {1997})}\BibitemShut {NoStop}%
\bibitem [{\citenamefont {Quinn}\ and\ \citenamefont
  {Wald}(1997)}]{QuinWald97}%
  \BibitemOpen
  \bibfield  {author} {\bibinfo {author} {\bibfnamefont {T.~C.}\ \bibnamefont
  {Quinn}}\ and\ \bibinfo {author} {\bibfnamefont {R.~M.}\ \bibnamefont
  {Wald}},\ }\href {\doibase 10.1103/PhysRevD.56.3381} {\bibfield  {journal}
  {\bibinfo  {journal} {Phys. Rev. D}\ }\textbf {\bibinfo {volume} {56}},\
  \bibinfo {pages} {3381} (\bibinfo {year} {1997})}\BibitemShut {NoStop}%
\bibitem [{\citenamefont {Detweiler}\ and\ \citenamefont
  {Whiting}(2003)}]{DetwWhit03}%
  \BibitemOpen
  \bibfield  {author} {\bibinfo {author} {\bibfnamefont {S.~L.}\ \bibnamefont
  {Detweiler}}\ and\ \bibinfo {author} {\bibfnamefont {B.~F.}\ \bibnamefont
  {Whiting}},\ }\href {\doibase 10.1103/PhysRevD.67.024025} {\bibfield
  {journal} {\bibinfo  {journal} {Phys. Rev. D}\ }\textbf {\bibinfo {volume}
  {67}},\ \bibinfo {pages} {024025} (\bibinfo {year} {2003})}\BibitemShut
  {NoStop}%
\bibitem [{\citenamefont {Poisson}\ \emph {et~al.}(2011)\citenamefont
  {Poisson}, \citenamefont {Pound},\ and\ \citenamefont {Vega}}]{Pois11}%
  \BibitemOpen
  \bibfield  {author} {\bibinfo {author} {\bibfnamefont {E.}~\bibnamefont
  {Poisson}}, \bibinfo {author} {\bibfnamefont {A.}~\bibnamefont {Pound}}, \
  and\ \bibinfo {author} {\bibfnamefont {I.}~\bibnamefont {Vega}},\ }\href
  {\doibase 10.12942/lrr-2011-7} {\bibfield  {journal} {\bibinfo  {journal}
  {Living Reviews in Relativity}\ }\textbf {\bibinfo {volume} {14}},\ \bibinfo
  {pages} {7} (\bibinfo {year} {2011})}\BibitemShut {NoStop}%
\bibitem [{\citenamefont {Barack}(2000)}]{Bara00}%
  \BibitemOpen
  \bibfield  {author} {\bibinfo {author} {\bibfnamefont {L.}~\bibnamefont
  {Barack}},\ }\href {\doibase 10.1103/PhysRevD.62.084027} {\bibfield
  {journal} {\bibinfo  {journal} {Phys. Rev. D}\ }\textbf {\bibinfo {volume}
  {62}},\ \bibinfo {pages} {084027} (\bibinfo {year} {2000})}\BibitemShut
  {NoStop}%
\bibitem [{\citenamefont {Burko}\ and\ \citenamefont {Liu}(2001)}]{Burk01}%
  \BibitemOpen
  \bibfield  {author} {\bibinfo {author} {\bibfnamefont {L.~M.}\ \bibnamefont
  {Burko}}\ and\ \bibinfo {author} {\bibfnamefont {Y.~T.}\ \bibnamefont
  {Liu}},\ }\href {\doibase 10.1103/PhysRevD.64.024006} {\bibfield  {journal}
  {\bibinfo  {journal} {Phys. Rev. D}\ }\textbf {\bibinfo {volume} {64}},\
  \bibinfo {pages} {024006} (\bibinfo {year} {2001})}\BibitemShut {NoStop}%
\bibitem [{\citenamefont {Zimmerman}\ and\ \citenamefont
  {Poisson}(2014)}]{ZimmPois14}%
  \BibitemOpen
  \bibfield  {author} {\bibinfo {author} {\bibfnamefont {P.}~\bibnamefont
  {Zimmerman}}\ and\ \bibinfo {author} {\bibfnamefont {E.}~\bibnamefont
  {Poisson}},\ }\href {\doibase 10.1103/PhysRevD.90.084030} {\bibfield
  {journal} {\bibinfo  {journal} {Phys. Rev. D}\ }\textbf {\bibinfo {volume}
  {90}},\ \bibinfo {pages} {084030} (\bibinfo {year} {2014})}\BibitemShut
  {NoStop}%
\bibitem [{\citenamefont {Linz}\ \emph {et~al.}(2014)\citenamefont {Linz},
  \citenamefont {Friedman},\ and\ \citenamefont {Wiseman}}]{Linz14}%
  \BibitemOpen
  \bibfield  {author} {\bibinfo {author} {\bibfnamefont {T.~M.}\ \bibnamefont
  {Linz}}, \bibinfo {author} {\bibfnamefont {J.~L.}\ \bibnamefont {Friedman}},
  \ and\ \bibinfo {author} {\bibfnamefont {A.~G.}\ \bibnamefont {Wiseman}},\
  }\href {\doibase 10.1103/PhysRevD.90.084031} {\bibfield  {journal} {\bibinfo
  {journal} {Phys. Rev. D}\ }\textbf {\bibinfo {volume} {90}},\ \bibinfo
  {pages} {084031} (\bibinfo {year} {2014})}\BibitemShut {NoStop}%
\bibitem [{\citenamefont {Zimmerman}(2015)}]{Zimm15}%
  \BibitemOpen
  \bibfield  {author} {\bibinfo {author} {\bibfnamefont {P.}~\bibnamefont
  {Zimmerman}},\ }\href {\doibase 10.1103/PhysRevD.92.064040} {\bibfield
  {journal} {\bibinfo  {journal} {Phys. Rev. D}\ }\textbf {\bibinfo {volume}
  {92}},\ \bibinfo {pages} {064040} (\bibinfo {year} {2015})}\BibitemShut
  {NoStop}%
\bibitem [{\citenamefont {Pound}\ \emph {et~al.}(2005)\citenamefont {Pound},
  \citenamefont {Poisson},\ and\ \citenamefont {Nickel}}]{Poun05}%
  \BibitemOpen
  \bibfield  {author} {\bibinfo {author} {\bibfnamefont {A.}~\bibnamefont
  {Pound}}, \bibinfo {author} {\bibfnamefont {E.}~\bibnamefont {Poisson}}, \
  and\ \bibinfo {author} {\bibfnamefont {B.~G.}\ \bibnamefont {Nickel}},\
  }\href {\doibase 10.1103/PhysRevD.72.124001} {\bibfield  {journal} {\bibinfo
  {journal} {Phys. Rev. D}\ }\textbf {\bibinfo {volume} {72}},\ \bibinfo
  {pages} {124001} (\bibinfo {year} {2005})}\BibitemShut {NoStop}%
\bibitem [{\citenamefont {Hinderer}\ and\ \citenamefont
  {Flanagan}(2008)}]{Hind08}%
  \BibitemOpen
  \bibfield  {author} {\bibinfo {author} {\bibfnamefont {T.}~\bibnamefont
  {Hinderer}}\ and\ \bibinfo {author} {\bibfnamefont {E.~E.}\ \bibnamefont
  {Flanagan}},\ }\href {\doibase 10.1103/PhysRevD.78.064028} {\bibfield
  {journal} {\bibinfo  {journal} {Phys. Rev. D}\ }\textbf {\bibinfo {volume}
  {78}},\ \bibinfo {pages} {064028} (\bibinfo {year} {2008})}\BibitemShut
  {NoStop}%
\bibitem [{\citenamefont {{Hopman}}\ and\ \citenamefont
  {{Alexander}}(2005)}]{Hopm05}%
  \BibitemOpen
  \bibfield  {author} {\bibinfo {author} {\bibfnamefont {C.}~\bibnamefont
  {{Hopman}}}\ and\ \bibinfo {author} {\bibfnamefont {T.}~\bibnamefont
  {{Alexander}}},\ }\href {\doibase 10.1086/431475} {\bibfield  {journal}
  {\bibinfo  {journal} {Astrophysical Journal}\ }\textbf {\bibinfo {volume}
  {629}},\ \bibinfo {pages} {362} (\bibinfo {year} {2005})}\BibitemShut
  {NoStop}%
\bibitem [{\citenamefont {Chandrasekhar}(1992)}]{Chan92}%
  \BibitemOpen
  \bibfield  {author} {\bibinfo {author} {\bibfnamefont {S.}~\bibnamefont
  {Chandrasekhar}},\ }\enquote {\bibinfo {title} {The mathematical theory of
  black holes},}\ \ (\bibinfo  {publisher} {Oxford University Press},\ \bibinfo
  {year} {1992})\ Chap.~\bibinfo {chapter} {5}, pp.\ \bibinfo {pages}
  {215--224}\BibitemShut {NoStop}%
\bibitem [{\citenamefont {Sorce}\ and\ \citenamefont
  {Wald}(2017)}]{SorcWald17}%
  \BibitemOpen
  \bibfield  {author} {\bibinfo {author} {\bibfnamefont {J.}~\bibnamefont
  {Sorce}}\ and\ \bibinfo {author} {\bibfnamefont {R.~M.}\ \bibnamefont
  {Wald}},\ }\href {\doibase 10.1103/PhysRevD.96.104014} {\bibfield  {journal}
  {\bibinfo  {journal} {Phys. Rev. D}\ }\textbf {\bibinfo {volume} {96}},\
  \bibinfo {pages} {104014} (\bibinfo {year} {2017})}\BibitemShut {NoStop}%
\bibitem [{\citenamefont {Zimmerman}\ \emph {et~al.}(2013)\citenamefont
  {Zimmerman}, \citenamefont {Vega}, \citenamefont {Poisson},\ and\
  \citenamefont {Haas}}]{ZimmETC13}%
  \BibitemOpen
  \bibfield  {author} {\bibinfo {author} {\bibfnamefont {P.}~\bibnamefont
  {Zimmerman}}, \bibinfo {author} {\bibfnamefont {I.}~\bibnamefont {Vega}},
  \bibinfo {author} {\bibfnamefont {E.}~\bibnamefont {Poisson}}, \ and\
  \bibinfo {author} {\bibfnamefont {R.}~\bibnamefont {Haas}},\ }\href {\doibase
  10.1103/PhysRevD.87.041501} {\bibfield  {journal} {\bibinfo  {journal} {Phys.
  Rev. D}\ }\textbf {\bibinfo {volume} {87}},\ \bibinfo {pages} {041501}
  (\bibinfo {year} {2013})}\BibitemShut {NoStop}%
\bibitem [{\citenamefont {{Martel}}\ and\ \citenamefont
  {{Poisson}}(2005)}]{MartPois05}%
  \BibitemOpen
  \bibfield  {author} {\bibinfo {author} {\bibfnamefont {K.}~\bibnamefont
  {{Martel}}}\ and\ \bibinfo {author} {\bibfnamefont {E.}~\bibnamefont
  {{Poisson}}},\ }\href {\doibase 10.1103/PhysRevD.71.104003} {\bibfield
  {journal} {\bibinfo  {journal} {Phys. Rev. D}\ }\textbf {\bibinfo {volume}
  {71}},\ \bibinfo {pages} {104003} (\bibinfo {year} {2005})}\BibitemShut
  {NoStop}%
\end{thebibliography}%

\end{document}